%% file: tvlsi.tex
\documentclass[journal,10pt]{IEEEtran}
\input{./ieee-compact}

\graphicspath{{./figs/}}

\newcommand{\revise}[1]{\textcolor{black}{#1}}

\begin{document}

\newtheorem{property}{\textbf{Property}}
\newtheorem{definition}{\textbf{Definition}}
\newtheorem{lemma}{\textbf{Lemma}}
\newtheorem{theorem}{\textbf{Theorem}}
\newtheorem{corollary}{\textbf{Corollary}}


\title{Integrated Optimization of Partitioning, Scheduling, and Floorplanning for Partially Dynamically Reconfigurable Systems}

\author{
Song Chen,~\IEEEmembership{Member,~IEEE},
Jinglei Huang,
Xiaodong Xu,
Bo Ding,
and Qi Xu

\thanks{This work was partially supported by the National Natural Science Foundation of China (NSFC) under grant No.\revise{61874102}, No.61732020  and 61674133.
The authors would like to thank the Information Science Laboratory Center of USTC for the hardware \& software services.
}
\thanks{
S.~Chen, X.~Xu, and B. Ding are with the Department of Electronic Science and Technology, University of Science and Technology of China, China (email: songch@ustc.edu.cn;xxd0210@mail.ustc.edu.cn;dingbo@mail.ustc.edu.cn).

J.~Huang is with State Key Laboratory of Air Traffic Management System and Technology, China (email:huangjl@mail.ustc.edu.cn). 

Q.~Xu is with the School of Electronic Science and Applied Physics, Hefei University of Technology, China (email: xuqi@hfut.edu.cn).}
}

\markboth{IEEE TRANSACTIONS ON COMPUTER-AIDED DESIGN OF INTEGRATED CIRCUITS AND SYSTEMS, VOL., NO.}
{S. Chen \MakeLowercase{\textit{et al.}}: \textsc{An Integrated Optimization Framework for Partitioning, Scheduling, and Floorplanning on Partially Dynamically Reconfigurable FPGAs}}

\maketitle

\begin{abstract}
Confronted with the challenge of high performance for applications and the restriction of hardware resources for field-programmable gate arrays (FPGAs), partial dynamic reconfiguration (PDR) technology is anticipated to accelerate the reconfiguration process and alleviate the device shortage.
In this paper, we propose an integrated optimization framework for task partitioning, scheduling and floorplanning on partially dynamically reconfigurable FPGAs.
The partition, schedule, and floorplan of the tasks are represented by the partitioned sequence triple $P$-$ST$ $(PS, QS, RS)$, where $(PS, QS)$ is a hybrid nested sequence pair ($HNSP$) for representing the spatial and temporal partitions, as well as the floorplan, and $RS$ is the partitioned dynamic configuration order of the tasks.
The floorplanning and scheduling of task modules can be computed from the partitioned sequence triple $P$-$ST$ in $O(n^2)$ time.
To integrate the exploration of the scheduling and floorplanning design space, we use a simulated annealing-based search engine and elaborate a perturbation method, where a randomly chosen task module is removed from the partition sequence triple and then re-inserted into a proper position selected from all the {$O(n^3)$} possible combinations of partition, schedule and floorplan.
We also prove a sufficient and necessary condition for the feasibility of the partitioning of tasks and scheduling of task configurations, and derive conditions for the feasibility of the insertion points in a P-ST.
The experimental results demonstrate the efficiency and effectiveness of the proposed framework.
\end{abstract}
\begin{IEEEkeywords}
Partitioning, scheduling, floorplanning, partially dynamically reconfigurable, FPGAs, partitioned sequence triple
\end{IEEEkeywords}

\section{Introduction}
In recent decades, reconfigurable hardware, and field programmable gate arrays (FPGAs) in particular, have received much attention because of their ability to be reconfigured to any custom desired computing architecture rapidly~\cite{tessier2015reconfigurable}.
We can construct an entire hardware system on an FPGA chip or include an FPGA on a system-on-chip to provide hardware programmability. Traditionally, FPGAs are exploited using compile-time (static) reconfiguration, and the configuration remains the same throughout the running time of an application.
To change the configuration, we have to stop the computation, reconfigure the chip by means of power-on resetting, and then start a new application.
With the evolution of FPGA technology, dynamic reconfiguration (DR) has been developed, which provides more flexibility to reconfigure the FPGA by changing its predetermined functions at run-time. Through DR, one large application can be partitioned into smaller tasks; then, the tasks can be sequentially configured at run-time. In this process, the entire chip must be reconfigured for each task; thus, significant reconfiguration overhead is incurred for loading the configuration each time~\cite{hauck2010reconfigurable}.

To reduce the reconfiguration overhead and improve performance, several techniques are employed in modern FPGA architectures, such as partially dynamic reconfiguration (PDR), module reuse, and configuration prefetching, where PDR is a technique that reconfigures part of the FPGA at run-time while retaining normal operation of the remaining areas of the FPGA~\cite{koch2012partial}.
By applying the PDR technique, different tasks can be executed and configured in parallel, and a portion of the configuration latency can be hidden by careful scheduling of the configurations and executions of tasks. Hereafter, the FPGA, with the characteristic of PDR, is regarded as a partially dynamically reconfigurable FPGA (PDR-FPGA).

To implement a large application composed of task modules on a PDR-FPGA,
we must consider two problems: when the task modules should be configured and executed and where the task modules should be placed. The former is a scheduling problem, and the latter is a floorplanning problem.
Unfortunately, both of them are NP-hard \cite{murata1995rectangle}~\cite{gary1979computers}.
In addition, to enable PDR, the reconfigurable resources on the FPGA are partitioned into several reconfigurable regions, which will be dynamically reconfigured to realize different tasks over time.
Therefore, the number of partitioned reconfigurable regions and their sizes should be considered in this process.

\subsection{Related Work}
Many studies have focused on partitioning, scheduling, and floorplanning for PDR.
R. Cordone et al. \cite{cordone2009partitioning} proposed an integer linear programming (ILP) based method and a heuristic method for partitioning and scheduling task graphs on PDR-FPGAs, where configuration prefetching and module reuse are considered to minimize the reconfiguration overhead.
A. Purgato et al. \cite{purgato2016resource} proposed a fast task scheduling heuristic to schedule the tasks in either the hardware or the software with minimization of the overall execution time on partially reconfigurable systems. 
However, the proposed method only focuses on generating reconfigurable regions to satisfy the resource requirements, which will easily cause the final result to fail to produce a valid floorplan.
Y. Jiang et al. \cite{jiang2007temporal} proposed a network flow-based multi-way task partitioning algorithm to minimize the total communication costs across temporal partitions. However, in this work, the partitioning is simplified
without considering the partial reconfiguration, and it is difficult to effectively estimate the communication costs without the floorplan information.
All the aforementioned works mainly focus on partitioning/scheduling of the tasks without consideration of the floorplan, which will often cause the schedule to fail to be floorplanned effectively, as they do not consider the resource constraints on the FPGA chips.

E. A. Deiana et al. \cite{deiana2015multiobjective} proposed a mixed-integer linear programming (MILP) based scheduler for mapping and scheduling applications on partially reconfigurable FPGAs, and if the schedule cannot be successfully floorplanned, the scheduler is re-executed until a feasible floorplan is identified.
However, the time-consuming MILP based method is impractical for large applications. 
In addition, scheduling and floorplanning are solved separately, which can cause large communication costs in the spatial domain.
M. Vasilko \cite{vasilko1999dynasty} proposed a temporal floorplanning method for solving the scheduling and floorplanning of dynamically reconfigurable systems.
P. Yuh et al. \cite{yuh2004temporal} \cite{yuh2007temporal} modeled the tasks as three-dimensional (3D) boxes
and proposed simulated annealing-based 3D floorplanners 
to solve the floorplanning and scheduling problems of the tasks.
However, the task modules are assumed to be reconfigured at any time and in any region, which may not match practical reconfigurable architectures. For example, in the Virtex 7 series FPGA chips from Xilinx \cite{vivado-pr-ug2017}, the reconfiguration partitions (dynamically reconfigurable regions) cannot be overlapped.
Given scheduled task graphs, many works have focused on the floorplanning of partially reconfigurable designs
\cite{montone2010placement,singhal2006multi,banerjee2011floorplanning,nan2013resource,rabozzi2014floorplanning,rabozzi2015floorplanning}.

The design of reconfigurable systems with PDR generally involves partitioning, scheduling, and floorplanning of the tasks, which are interdependent considering communication costs and system performance.
Therefore, these three problems have to be solved in an integrated optimization framework to effectively explore the design space.
However, the aforementioned works either solve the three problems sequentially, where, at most, a simple iterative refinement between scheduling and floorplanning is included, or solve only two of the three problems in an integrated framework.

\subsection{Contributions}
\label{sec:contributions}
In this paper, we propose an integrated optimization framework for task partitioning, scheduling, and floorplanning on partially dynamically reconfigurable FPGAs. This paper expands our previous work \cite{xu2017glsvlsi}. Numerous theoretical analyses are provided for the feasibility of the $P$-$ST$s (defined below). The main contributions of this paper are outlined as follows.

1). The term $P$-$ST$ $(PS, QS, RS)$ is proposed to represent the partitions, schedule, and floorplan of $n$ task modules, where $PS$, $QS$, and $RS$ are the sequences of $n$ task modules.
$(PS,$ $QS)$ is regarded as a hybrid nested sequence pair ($HNSP$) representing the floorplan with spatial and temporal partition, and $RS$ is the partitioned dynamic configuration order of the tasks.
The floorplan can be computed from the $HNSP$ in $O(nloglogn)$ time, and the schedule of tasks can be computed in $O(n^2)$ time by solving a single-source longest-path problem on a reconfiguration constraint graph ($RCG$), which is constructed based on $P$-$ST$ and the task precedence graph.

2). We elaborate a perturbation method to integrate the exploration of the schedule and floorplan design space into simulated annealing-based searching.
In the perturbation, a randomly chosen task module is removed from a $P$-$ST$ and is then re-inserted into the partitioned sequence triple at a proper position selected from all {$O(n^3)$} possible insertion points, which are efficiently evaluated in $O(n^{4})$ time based on an \textit{insertion point} enumeration procedure.

3). We prove a sufficient and necessary condition for the feasibility of the partitioning of tasks and scheduling of task configurations, which is not included in \cite{xu2017glsvlsi}, and derive conditions for the feasibility of the insertion points in a $P$-$ST$.

The experimental results demonstrate the efficiency and effectiveness of the proposed optimization framework.

The remainder of the paper is organized as follows. Section~\ref{sec:problem} describes the target hardware architecture and the problem definition.
Section~\ref{sec:ST} discusses the representation of a sequence triple.
Section~\ref{sec:perturbation} shows the optimization framework to explore the design space of partitioning, scheduling and floorplanning of task modules.
Experimental results and conclusions are shown and discussed in Section~\ref{sec:experiments} and Section~\ref{sec:conclusion}, respectively.

\section{Problem Description}
\label{sec:problem}
\subsection{Dynamically Reconfigurable Architecture}\label{sec:arch}
The dynamically reconfigurable system typically includes a host processor, an FPGA chip, an external memory, and the communication infrastructure among them.
The host processor and communication infrastructure could be on-chip or off-chip.
Pre-synthesized task modules are stored in off-chip external memory in the form of bitstreams.
According to the scheduled sequence and floorplanned locations, the host processor deploys task modules on the FPGAs.

Modern FPGAs have evolved into complex heterogeneous and hierarchical devices.
However, the basic logic cell still comprises configurable logic blocks (CLBs) \cite{chen2006fpga}.
In the target architecture, the CLB is the smallest reconfigurable element.
Configuration bitstreams are transferred into FPGAs using one configuration port, which is an external Joint Test Action Group protocol or an internal configuration access port (ICAP).

On the other hand, PDR is subject to the technology limitation, which is that the configuration process of a task module must not disrupt the execution of other task modules \cite{montone2010placement}. Thus, generally, dynamically reconfigurable regions (DRRs), where the task modules are dynamically reconfigured in a manner similar to that of a context (time layer) switching mode, are used for implementing partial reconfiguration. On an FPGA chip, we can have multiple DRRs and one DRR can be dynamically reconfigured while the others continue to execute.

A DRR is a rectangular region on FPGAs because irregular-shaped reconfiguration regions (such as T or L shapes) can introduce routing restriction issues \cite{vivado-pr-ug2017}. A task can be implemented as a rectangular hardware module on the FPGA. The module area represents the occupied CLBs (the number of rows and columns on the FPGA).

\subsection{Problem Definition}
\label{sec:problem-def}
The design is composed of pre-synthesized tasks whose resource usage and internal routing are predetermined.
Let $M\!=\!\{m_i|1\!\le\!i\!\le\!n\}$ be a set of $n$ tasks. 
A task $m_i$, has a physical attribute vector, $(w_i, h_i, c_i, t_i)$. \revise{The meanings are shown in Table \ref{tab:notations}}. \revise{$c_i$ is proportional to the area and is estimated by $c_{clb}\times w_i \times h_i$, where $c_{clb}$ is the configuration time of a single CLB.}
\begin{table}[!tb]
\renewcommand{\arraystretch}{1.2}
\centering
\caption{\revise{Some frequently used notations.}}
\label{tab:notations}
    \begin{tabular}{|c|p{6.8cm}|}
    \hline
    $m_i$        & $1\le i \le n$ and $m_i$ is a task module.           \\\hline
    $w_i,h_i$    & number of CLB rows and CLB columns required by $m_i$. \\\hline
    $c_i,t_i$    & configuration span (time) and execution span (time) of $m_i$. \\\hline
    $bc_i,bt_i$  & start configuration time /start execution time of $m_i$. \\\hline
    $drr_i$      & $1\le i \le N$ and $drr_i$ is a dynamically reconfigurable region. \\\hline
    $drr(m_i)$   & the DRR where $m_i$ is located. \\\hline
    $tl^i_j$     & the $j$-th time layer in $drr_i$.\\\hline
    $tl(m_i)$    & the time layer where $m_i$ is located. \\\hline
    $c_{tl^i_j}$ & configuration span of time layer $tl^i_j$. \\\hline
    $bc_{tl^i_j}$& start configuration time of time layer $tl^i_j$. \\\hline
    $CO[tl^i_j]$ & configuration order of time layer $tl^i_j$. \\\hline
    $lt\_s^i_j$  & start of lifetime of a time layer $tl^i_j$. \\\hline
    $lt\_e^i_j$  & end of lifetime of a time layer $tl^i_j$. \\\hline
    $LT[tl^i_j]$ & =($lt\_s^i_j$, $lt\_e^i_j$), lifetime of a time layer $tl^i_j$.\\\hline
\end{tabular}
\end{table}

The data dependencies among these tasks are given as a task dependence graph, $TG=(V_{TG},E_{TG})$, where $V_{TG}=M$ and $E_{TG}=\{(m_i,m_j )|1\le i,j\le n,i \ne j$, and $m_i$ must end before $m_j$ starts$\}$.
$TG'=(V_{TG}, E_{TG'})$ denotes the transitive closure of $TG$.

The partitioning, scheduling, and floorplanning of PDR are formulated as follows:

In the spatial domain, the $n$ tasks are partitioned into DRRs.
Let $N$ be the number of DRRs.
\begin{definition}
	The DRRs are denoted as $DRR=\{drr_i|1\le i\le N, drr_i\subseteq M, \bigcup_{i=1}^N drr_i=M\}$, where $\forall i\ne j$, $drr_i\cap drr_j=\emptyset$.
	If $m_k\in drr_i$, we denote the DRR of $m_k$ as $drr(m_k) = drr_i$.
\end{definition}

In the temporal domain, the $n$ tasks are partitioned into different time layers to reuse the resources of DRRs. A time layer is configured as a whole. Thus, in the same DRR, a time layer can only be configured after the completion of all the tasks in the previous time layer.
Let $l_i$ be the number of time layers in $drr_i$.
\begin{definition}
	The time layers are denoted as $TL=\{tl_j^i|\forall 1$ $\le i\le N, 1\le j\le l_i, tl_j^i\subseteq drr_i, \bigcup_{j=1}^{l_i} tl_j^i=drr_i\}$, where $\forall j_1\ne j_2$, $tl_{j_1}^i\cap tl_{j_2}^i=\emptyset$. If $m_k\in tl_j^i$, we denote the time layer of $m_k$ as $tl(m_k)$ and the total number of time layers as $|TL| = \sum_{i=1}^N l_i$.
\end{definition}

For convenience, we define $CO[tl^i_j]$ to be the configuration order of time layer $tl_j^i$, $1\le CO[tl^i_j]\le |TL|$ and stipulate that $CO[tl^i_j]$ $<$ $CO[tl^i_{j+1}]$, $1\le j < l_i$.

The configuration span  (time)  of the time layers in a DRR is {proportional} to the area of the DRR, we use $c_{drr_i}$ or $c_{tl_j^i}$ (= $c_{drr_i}$) to denote the configuration span of a time layer $tl_j^i$. To reduce the time complexity in the proposed integrated optimization framework, $c_{tl_j^i}$ is also under-estimated by summing the \revise{configuration time} of task modules: 
\begin{equation}
\label{equ:rough_ct}
c_{tl_j^i}=\!\sum\limits_{m_p\in tl_j^i} c_p
\end{equation}

For the scheduling, we consider the following constraints:
\begin{enumerate}
\item[(1)] The precedence constraints between tasks cannot be violated, that is,
$\forall (m_i,m_j)\in E_{TG}, bt_i+t_i\le bt_j$.
\item[(2)] A task must be configured before execution, that is,
$\forall 1\le i\le n, bc_{tl(m_i)}+c_{tl(m_i)}\le bt_i$.
\item[(3)] Considering the technical limitation of only one configuration port, the configuration span of time layers must be non-overlapped.
\item[(4)] In the same DRR, a time layer can only be configured after the execution of all the tasks in the previous time layer because they share the same hardware resources.
\end{enumerate}

The constraints for the floorplanning process are as follows.
\begin{enumerate}
\item[(5)] Each DRR occupies a rectangular region, and all the rectangular regions of the DRRs should be placed without overlapping each other and should be within the FPGA chip area, which is defined by the chip width and chip height (fixed-outline constraint).
\item[(6)] The task modules in the same time layer must be non-overlapped and placed within their corresponding DRR.
\end{enumerate}

Under the above constraints, we solve the \textit{partitioning} problem to determine $DRR$ and $TL$, the \textit{scheduling} problem to determine the start configuration time and start execution time of the tasks (time layers), and the \textit{flooprlanning} problem to determine the floorplan of DRRs and the floorplan of tasks inside the DRRs.

We define \textit{schedule length} to be the time from the beginning of the configuration process to the end of the executions of all tasks.
The objective is to find a reasonable floorplan of tasks on a PRD-FPGA while minimizing the \textit{schedule length} of designs as well as the communication costs among tasks.

\section{Partitioned Sequence Triple}\label{sec:ST}
\subsection{Representation}
\label{subsec:P-ST}
In this paper, a partitioned sequence triple ($P$-$ST$) is proposed to represent the partitioning, scheduling, and floorplanning of tasks for partially dynamically reconfigurable designs.

\begin{definition}
	\label{def:st}
	The partitioned sequence triple $P$-$ST$ is a 3-tuple of task sequences, $(PS, QS, RS)$, where $(PS, QS)$ forms a hybrid nested sequence pair ($HNSP$) to represent the spatial partition (DRR), the temporal partition (time layer) and the floorplan of the task modules, and $RS$ defines the configuration order of the time layers.
\end{definition}
{In a $P$-$ST$, task partitioning is constrained as follows:
\begin{enumerate}
\item[1)] The task modules in the same time layer will consecutively appear in $PS$, $QS$, and $RS$.
\item[2)] The task modules in the same DRR will consecutively appear in both $PS$ and $QS$.
\end{enumerate}
}

The structure of $P$-$ST$ is illustrated as follows:
\\
\noindent$(\langle\ldots\overbrace{[\ldots\underbrace{(\ldots m_p\ldots m_q\ldots)_{j_1}^i}_{tl_{j_1}^i}\ldots\underbrace{(\ldots m_r\ldots m_t\ldots)_{j_2}^i}_{tl_{j_2}^i}\ldots]_i}^{drr_i}\ldots\rangle$,
\noindent$\langle\ldots\overbrace{[\ldots\underbrace{(\ldots m_q\ldots m_p\ldots)_{j_1}^i}_{tl_{j_1}^i}\ldots\underbrace{(\ldots m_r\ldots m_t\ldots)_{j_2}^i}_{tl_{j_2}^i}\ldots]_i}^{drr_i}\ldots\rangle$,
\noindent$\langle\ldots\overbrace{\underbrace{(\ldots m_p\ldots m_q\ldots)_{j_1}^i}_{tl_{j_1}^i}}^{drr_i}\ldots\overbrace{\underbrace{(\ldots m_t\ldots m_r\ldots)_{j_2}^i}_{tl_{j_2}^i}}^{drr_i},\ldots\rangle)$.
\\

In a $P$-$ST$, $( \cdot )_j^i$ denotes the sequence of tasks in the time layer $tl_j^i$ and $[ \cdot ]_i$ is the sequence of tasks in the DRR $drr_i$.

\revise{An} $HNSP(PS, QS)$ imposes the position relationship between each pair of task modules as follows.
\begin{definition}
	\label{def:hnsp}
	if $tl(m_i)=tl(m_j)$ or $drr(m_i)\ne drr(m_j)$, then\\
	$( \langle\ldots m_{i} \ldots m_{j} \ldots\rangle, \langle\ldots m_{i} \ldots m_{j} \ldots\rangle )$ $\rightarrow$ $m_{i}$ is left to $m_{j}$;\\
	$( \langle\ldots m_{j} \ldots m_{i} \ldots\rangle, \langle\ldots m_{i} \ldots m_{j} \ldots\rangle )$ $\rightarrow$ $m_{i}$ is below $m_{j}$.
\end{definition}

Notice that the relationship between the task modules from different time layers in the same DRR is not defined, as there are no non-overlapping constraints involved.
Without loss of generality, we require the task modules in the same time layer to occur consecutively in $PS$ and $QS$ for clarity in representing the partitions of time layers and the floorplan of time layers.

The configuration order of a time layer can be represented by a configuration sequence $RS$, which is defined as follows:
\begin{definition}
	\label{def:rs}
	Given an $RS$ sequence, ($\langle\!\ldots\! m_{i}\!\ldots\! m_{j}\! \ldots\!\rangle$), the configuration constraints are defined as follows:\\
	1) if $tl(m_i)\!=\!tl(m_j)$, then $m_{i}$ and $m_{j}$ are configured simultaneously, along with the corresponding time layer;\\
	2) if $tl(m_i)\!\ne\!tl(m_j)$, then $tl(m_{i})$ is configured before $tl(m_{j})$, and the configuration order relationship is $CO[tl(m_{i})]\!<\!CO[tl(m_{j})]$.
\end{definition}

In the $RS$, the ordering of task modules within a time layer makes no sense because the time layer is configured as a whole.

For example, a task graph $TG$ with ten task modules is shown in Fig.~\ref{fig:TG} and a $P$-$ST$ in this example is given as follows:
\begin{figure}[tbp]
    \centering
    \includegraphics[width=0.32\textwidth,height=0.18\textwidth]{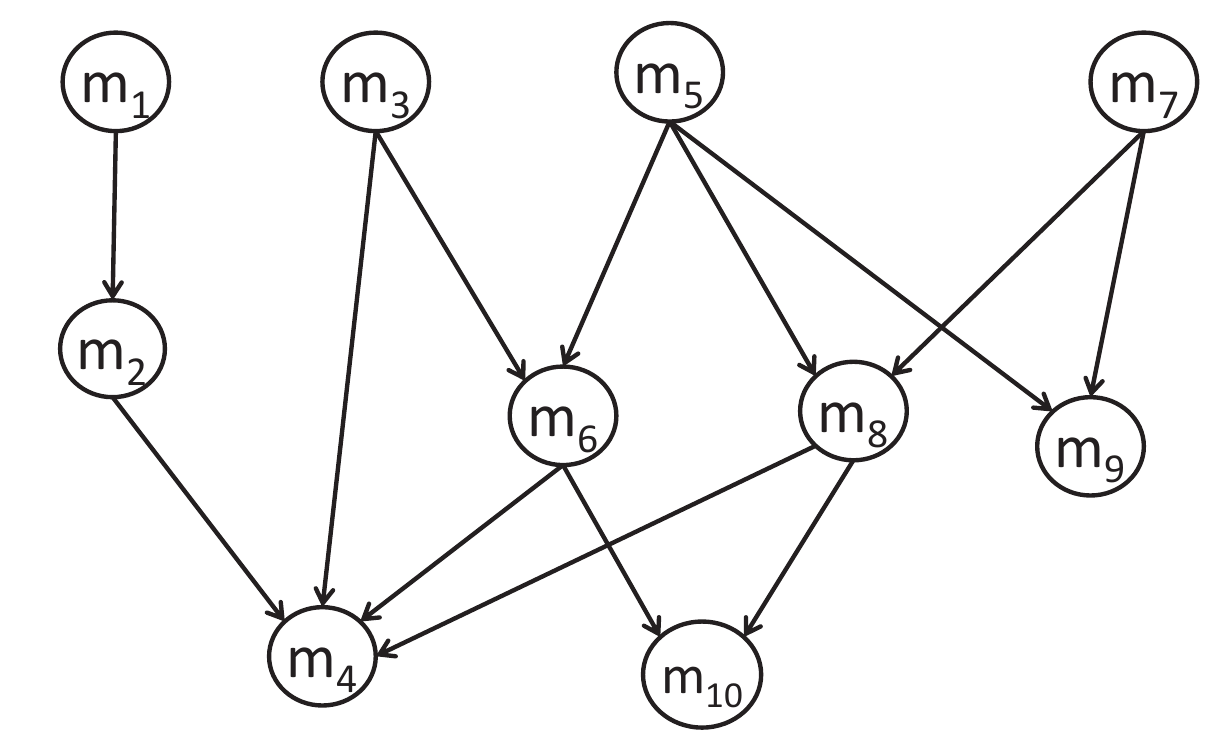}
    \caption{Task dependence graph $TG$ with ten task modules}
    \label{fig:TG}
\end{figure}
\begin{equation}
\label{equ:p-st-example}
\begin{split}
& (\langle[(1\ 2)_1^2\ (9\ 10)_2^2]_2\ [(8\ 7)_1^1]_1\ [(6)_1^4\ (4)_2^4]_4\ [(3)_1^3\ (5)_2^3]_3\rangle,\\
& \langle[(8\ 7)_1^1]_1\ [(2\ 1)_1^2\ (9\ 10)_2^2]_2\ [(3)_1^3\ (5)_2^3]_3\ [(6)_1^4\ (4)_2^4]_4\rangle,\\
& \langle(1\ 2)_1^2\ (3)_1^3\ (5)_2^3\ (6)_1^4\ (4)_2^4\ (7\ 8)_1^1\ (9\ 10)_2^2\rangle).
\end{split}
\end{equation}

\begin{figure}[htbp]
\small \centering
    \subfloat[Configuration order of time layers]{
        \includegraphics[width=0.40\textwidth,height=0.1\textwidth]{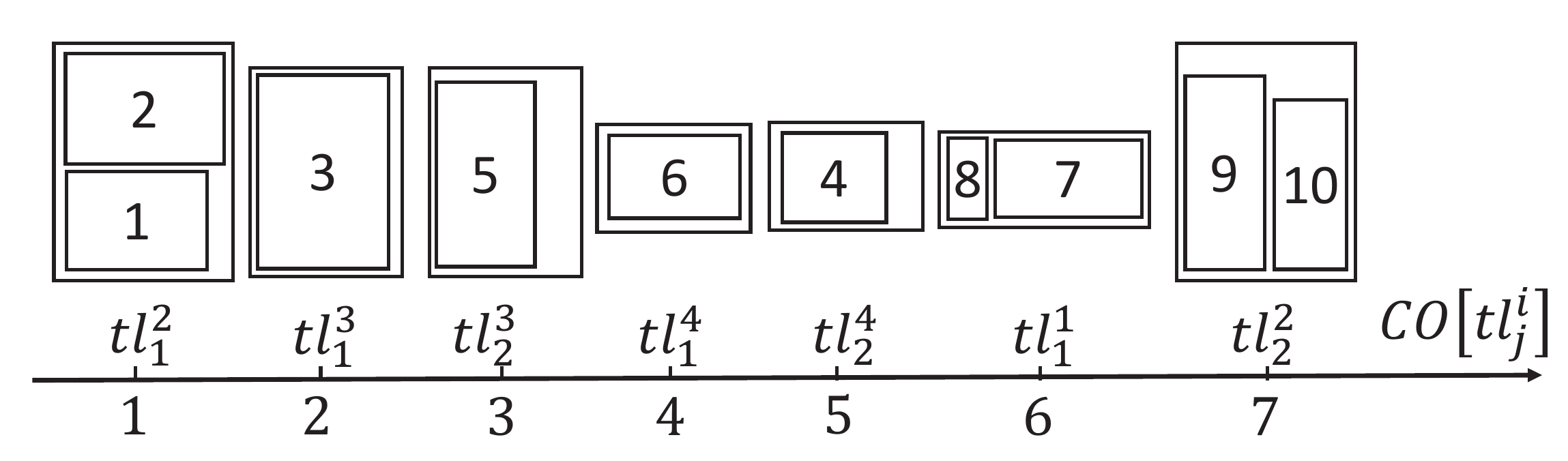}
        \label{fig:st-configuration}
    }
    \hspace{0.5em}
    \subfloat[Floorplan of DRRs and the tasks in the time layers]{
        \includegraphics[width=0.45\textwidth,height=0.25\textwidth]{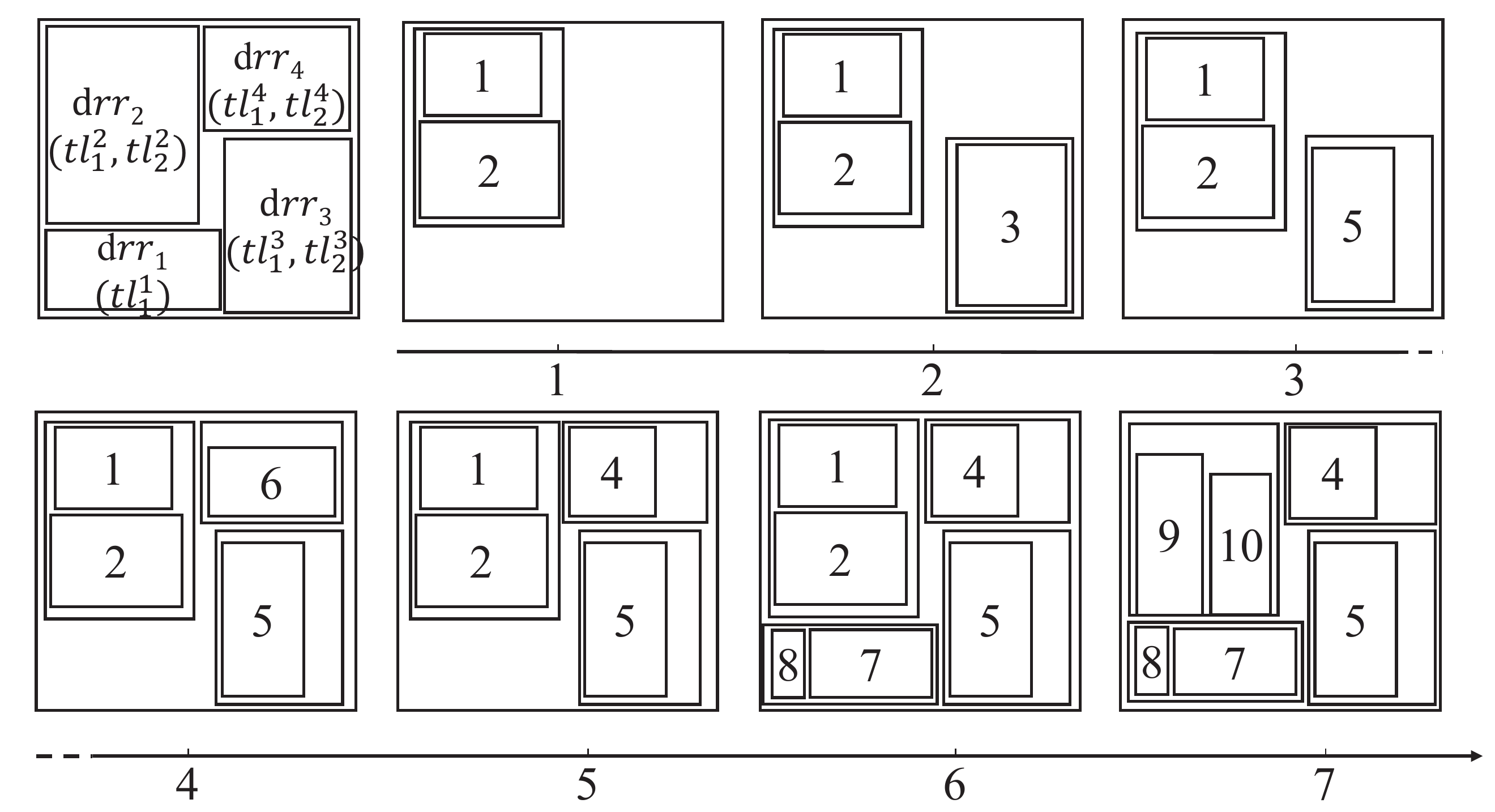}
        \label{fig:st-floorplan}
    }
\caption{From a $P$-$ST$ to the configuration order and the floorplan of DRRs and task modules in time layers.}
\label{fig:epST}
\end{figure}

For simplifying the notations, we use $i$ to represent the task module $m_i$ in the examples of $P$-$ST$. From the partitioned sequence triple $P$-$ST$, we can obtain the corresponding configuration order and floorplan on the FPGA as shown in Fig.~\ref{fig:epST}.

According to Definition~\ref{def:rs} and the given configuration sequence $RS\langle(1\ 2)_1^2\ (3)_1^3\ (5)_2^3\ (6)_1^4\ (4)_2^4\ (7\ 8)_1^1\ (9\ 10)_2^2\rangle$, Fig.~\ref{fig:st-configuration} shows the configuration order of the time layers.
First, the time layer $tl_{1}^{2}$ from $drr_2$ is configured and, second, the time layer $tl_{1}^{3}$ from $drr_3$ can be configured during the executions of $m_1$ and $m_2$.
The computation of the beginning configuration times of time layers will be discussed in Section \ref{sec:rcg}.

Considering the relationship between each pair of task modules defined in $Definition~\ref{def:hnsp}$, Fig.~\ref{fig:st-floorplan} shows the corresponding floorplan of task modules, where  $m_2$ is below  $m_1$ because they are in the same time layer ($tl(m_1)$ = $tl(m_2)$ = $tl_1^2$),
and $m_7$ is below $m_2$ because they are in different DRRs ($drr(m_7)=drr_1$ and $drr(m_2)=drr_2$).

Knowing the dimensions of task modules, we can compute the floorplan from the $HNSP$ in $O(nloglog(n))$ time by solving the longest weighted common subsequence of $PS$ and $QS$~\cite{tang2001fast} hierarchically. 
We can compute the floorplan of task modules within every DRR to obtain the occupied resource arrays of DRRs, and then compute the floorplan of DRRs to determine the total resource usage by regarding each DRR as a whole. The computation of the schedule will be discussed in the following subsections.

\subsection{Feasibility of Partition and Configuration Order}
\label{sec:feasibility}

Owing to the dependencies between tasks, not all the partitioned sequence triples $P$-$ST$ are feasible.
In this subsection, we prove a sufficient and necessary condition for the feasibility of partitions and configuration order of task modules.

\subsubsection{Lifetime of Time Layers}
The task modules in a time layer can be executed only after the configuration of the time layer and will be destroyed while configuring the next time layer (if more time layers exist) in the same DRR. Consequently, we have the following definition.
\begin{definition}
Given the spatial partition $DRR$, the temporal partition $TL$, and the configuration order of the time layers, we define the \textit{lifetime} of a time layer  $tl_j^i$, $LT[tl^i_j]=(lt\_s^i_j,\ lt\_e^i_j)$,  as follows.
$\forall tl^i_j\in drr_i,$
\begin{equation}
\label{equ:lifetime-def}
  \begin{split}
    &lt\_s^i_j = CO[tl^i_j],~and~lt\_e^i_j =
\begin{cases}
CO[tl^i_{j+1}], ~~1\le j < l_i;\\
\infty, ~~~~~~~~~j=l_i.
\end{cases}
  \end{split}
\end{equation}
\end{definition}
Note that the lifetime of a time layer is also the lifetime of the task modules in the time layer.

To discuss the feasibility of a configuration order, we define the dependencies between time layers based on the dependency graph of tasks given $DRR$ and $TL$.  A dependence graph \revise{$LTG(V_{LTG}, E_{LTG})$} is constructed as follows.

$V_{LTG} = TL$;
$E_{LTG} = \{(tl^{i_1}_{j_1}, tl^{i_2}_{j_2})|$
If there exist $m_{k_1}$ and $m_{k_2}$ respectively from $tl^{i_1}_{j_1}$ and $tl^{i_2}_{j_2}$ such that $(m_{k_1}, m_{k_2}) \in E_{TG}^{'}$\}. Note that $E_{TG}^{'}$ is the edge set of the transitive closure of $TG$.

\subsubsection{Dependencies Between Time Layers}
Given a configuration order, the dependencies between time layers fall into two groups: forward dependencies and backward dependencies.

\begin{definition}
\label{def:forward}
A dependence $(tl^{i_1}_{j_1}, tl^{i_2}_{j_2})\in E_{LTG}$ is \textit{forward} if $CO[tl^{i_1}_{j_1}] < CO[tl^{i_2}_{j_2}]$, which indicates that the output of a task module in a time layer, $tl^{i_1}_{j_1}$, is the input to a task module from a future time layer, $tl^{i_2}_{j_2}$.
\end{definition}

Forward dependencies are always feasible because even if the lifetime of a time layer ends, the computed data can be stored and used in the future.

\begin{definition}
A dependence $(tl^{i_1}_{j_1}, tl^{i_2}_{j_2})\in E_{LTG}$ is \textit{backward} if $CO[tl^{i_1}_{j_1}] > CO[tl^{i_2}_{j_2}]$, which indicates that the output of a task module in a time layer $tl^{i_1}_{j_1}$ is the input to a task module from an earlier configured time layer, $tl^{i_2}_{j_2}$.
\end{definition}

However, backward dependencies are infeasible if there is no overlapping between the lifetimes of the dependent time layers, $tl^{i_1}_{j_1}$ and $tl^{i_2}_{j_2}$, that is, $lt\_e^{i_2}_{j_2}<\ lt\_s^{i_1}_{j_1}$. In this situation, $tl^{i_2}_{j_2}$ is destroyed (replaced by a new time layer) before the time layer $tl^{i_1}_{j_1}$ is configured, so the input to a task module is generated after the task module has been destroyed.

\begin{figure}[htbp]
		\centering
       \includegraphics[width=0.48\textwidth]{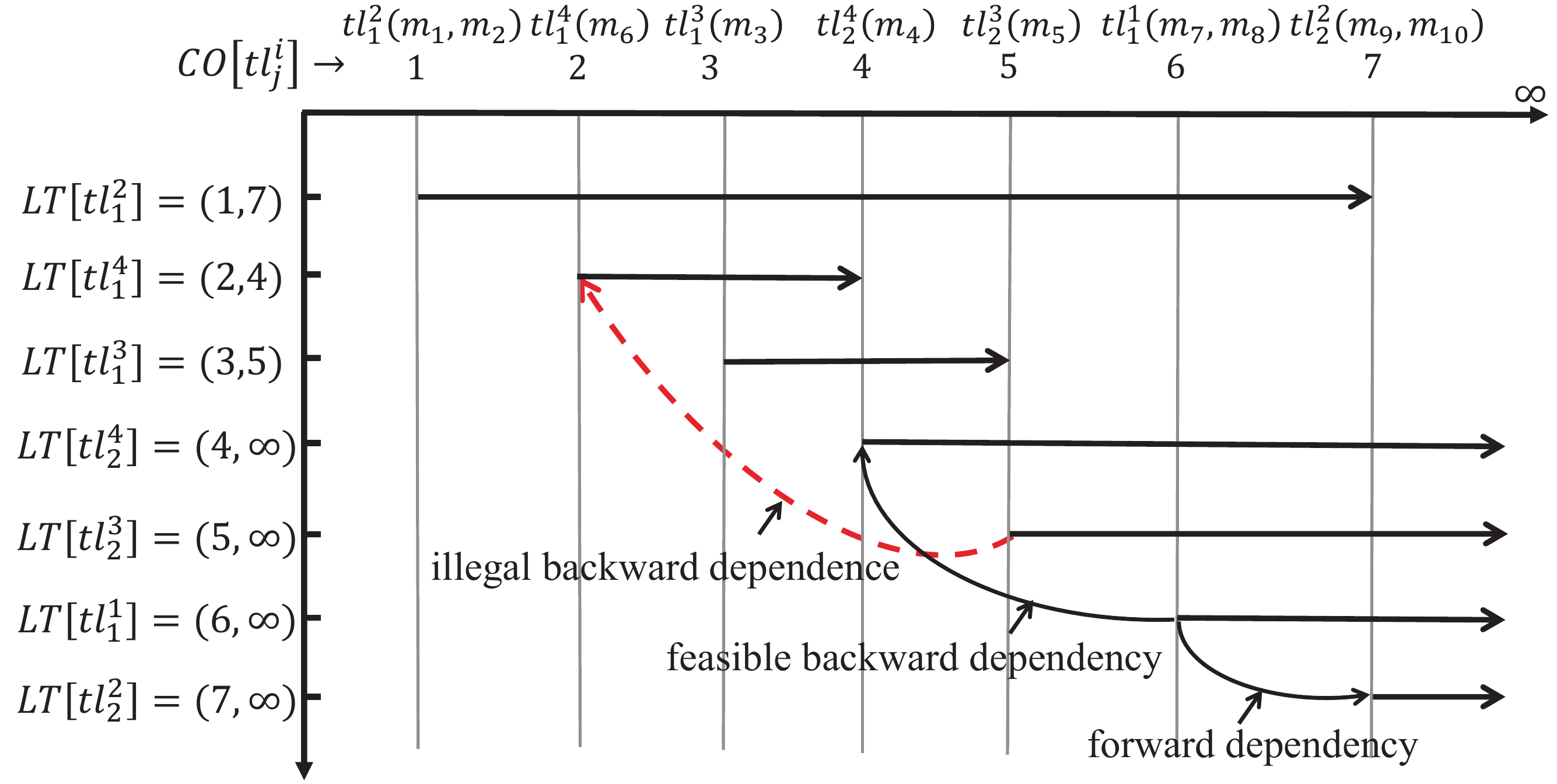}
		\caption{Lifetimes of time layers and an infeasible backward dependence.}
		\label{fig:feasibility.eps}
\end{figure}

Fig. \ref{fig:feasibility.eps} shows examples of lifetimes of time layers and the dependencies between time layers.
The spatial partition $DRR$ and the temporal partition $TL$ of the tasks are shown in Fig. \ref{fig:epST}, and the dependencies between tasks are shown in Fig. \ref{fig:TG}.
The configuration order of the time layers is as follows:
$
RS\langle(1\ 2)_1^2\  (6)_1^4\ (3)_1^3\ (4)_2^4\ (5)_2^3\  (7\ 8)_1^1\ (9\ 10)_2^2\rangle
$ 
(also shown as the x-axis in Fig.\ref{fig:feasibility.eps}).

The time layers $tl^4_1$ and $tl^3_2$ have backward dependence because $m_6$ needs the data from $m_5$, as shown in
Fig. \ref{fig:TG}, and their lifetimes $LT(tl^4_1)=(2,4)$ and $LT(tl^3_2)=(5,\infty)$ are non-overlapped. That is, $m_6$ in $tl^4_1$  is destroyed ( $tl^4_2$ in the same DRR $drr_4$ has occupied the hardware resource) before the execution of $m_5$ in $tl^3_2$.
Consequently, $m_6$ will never receive the data from $m_5$, so the configuration order of task modules shown in Fig. \ref{fig:feasibility.eps} is infeasible.

\subsubsection{Condition of Feasibility}
\label{subsec:sn-condition}

We thus argue that the given spatial partition, temporal partition, and configuration order is feasible if a schedule of executions and configurations of task modules can be computed without consideration of resource constraints.
We have the following theorem:
\begin{theorem}
  \label{the:feasibility}
The given spatial partition, temporal partition, and configuration order is feasible if and only if there are no backward dependencies between time layers that have no lifetime overlap.
\end{theorem}

\textbf{Proof.}
Given a partition, a configuration order, and the task dependency graph, we can construct a reconfiguration constraint graph ($RCG$) for scheduling the configurations of the time layers and the executions of the task modules, i.e., the computation of $bt_i$, $bc_i$, and $bc_{tl(m_{i})}$ defined in Section \ref{sec:problem-def}.

$RCG(V, E)$ is constructed by adding to the graph ($TG=(V, E)$) the vertex set $V_{LTG}$ and three edge sets representing the scheduling constraints.
$V_{RCG} = V_{TG}\cup V_{LTG}$, where $V_{LTG}$ represents time layers and is defined in Section \ref{sec:feasibility}.
$E_{RCG}=E_{TG}\cup E_{cr} \cup E_{ce} \cup E_{ec}$ and $E_{cr}, E_{ce}$, and $E_{ec}$ are defined as follows.

\begin{enumerate}
\item The set of edges represents the configuration order. $E_{cr} =\{ (tl^{i_1}_{j_1}, tl^{i_2}_{j_2})| CO[tl^{i_1}_{j_1}] < CO[tl^{i_2}_{j_2}]\}$.

\item
The set of edges indicates that a task $m_k$ must be executed only after the configuration of the time layer where $m_k$ is located. $E_{ce} = \{(tl^i_j, m_k)|tl^i_j\in V_{LTG}, m_k\in V_{TG}$ and $tl(m_k)=tl^i_j\}$.

\item The set of edges indicates that, in a DRR, a time layer \revise{must be} configured after the execution of all the tasks in the previous time layer because they share the same hardware resources. $E_{ec} = \{(m_k, tl^i_j)| tl^i_j\in V_{LTG}$, $m_k\in V_{TG}$,
and $tl(m_k)$ is the time layer  before $tl^i_j$ in $drr_i$ \}.
\end{enumerate}

A schedule can be computed only if the RCG is acyclic because a cycle produces a conflict in the constraints.

\textit{IF}. Here we show that if there are backward dependencies between the time layers that have no lifetime overlap, there will be a cycle in the RCG and hence the given partition and configuration order is infeasible.

\begin{figure}[htbp]
\small \centering
    \subfloat[Cycle formed by a backward dependence between time layers.]{
        \includegraphics[width=0.21\textwidth,height=0.17\textwidth]{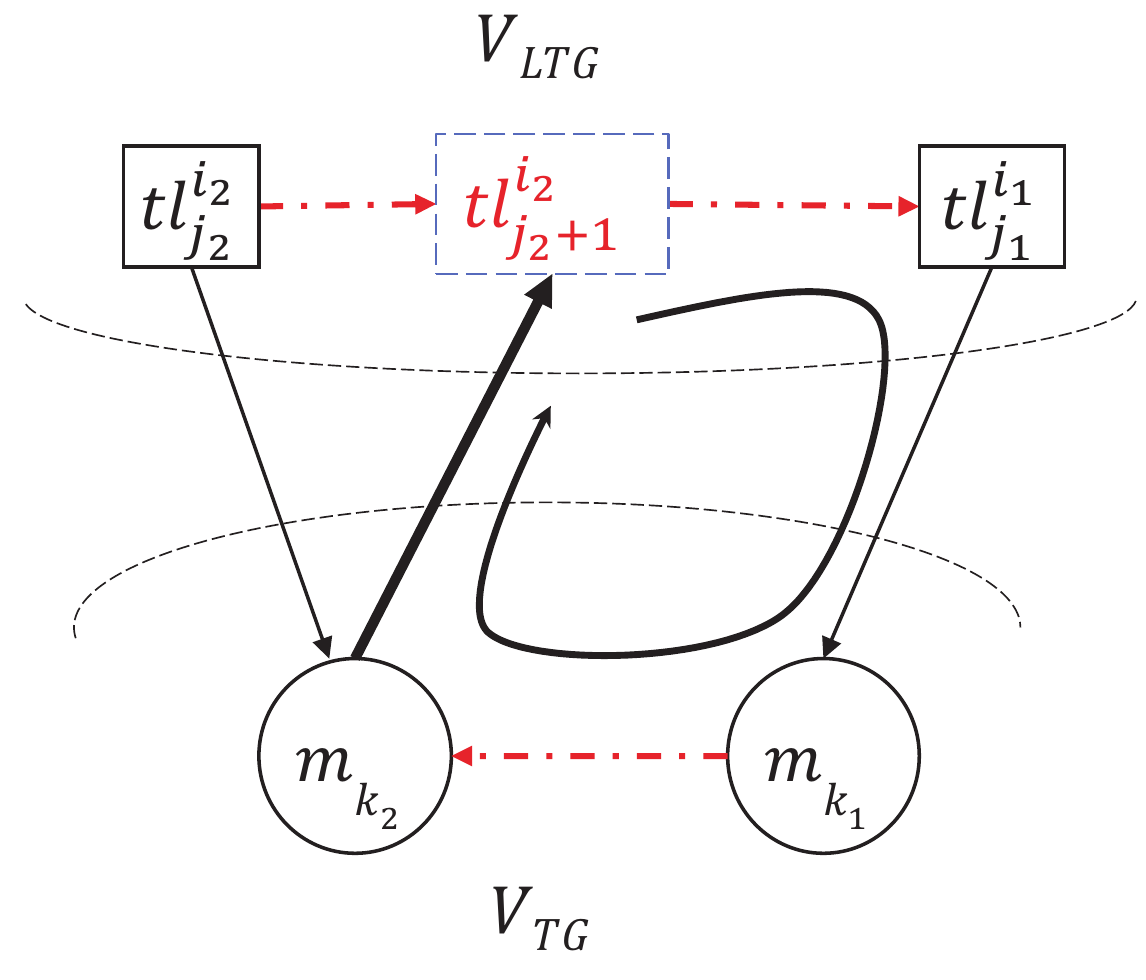}
        \label{fig:rcg-cycle-1}
    }
    \hspace{1.0em}
    \vspace{0.3em}
    \subfloat[The cycle causes an illegal backward dependence.]{
        \includegraphics[width=0.21\textwidth,height=0.17\textwidth]{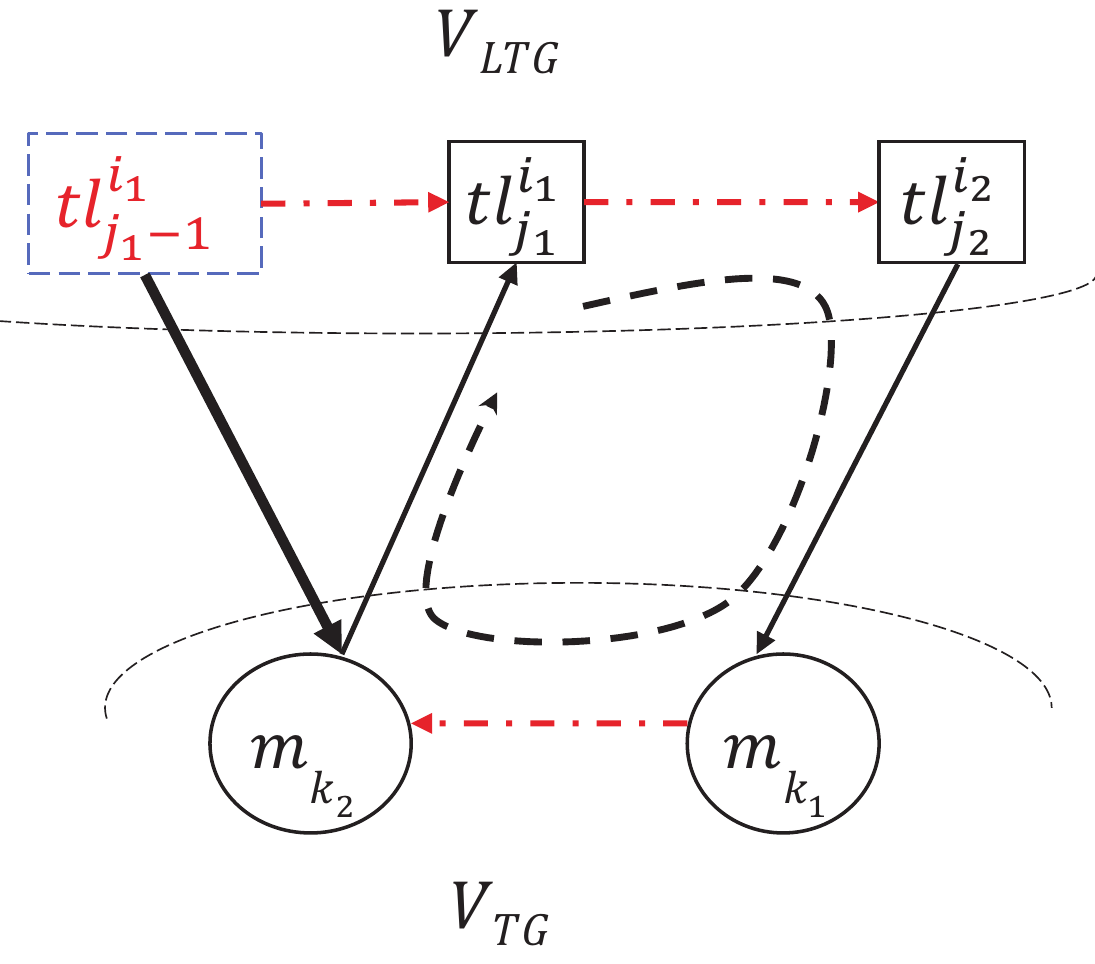}
        \label{fig:rcg-cycle-2}
    }
\caption{A cycle in RCG and the backward dependences.}
\label{fig:insertpoint}
\end{figure}

A pair of time layers, $tl^{i_1}_{j_1}$ and $tl^{i_2}_{j_2}$ with $ CO[tl^{i_1}_{j_1}] > CO[tl^{i_2}_{j_2}]$, have a backward dependence if there are two task modules, $m_{k_1}$ and $m_{k_2}$, respectively from $tl^{i_1}_{j_1}$ and $tl^{i_2}_{j_2}$ and there is a direct or indirect data dependence between them ($(m_{k_1}, m_{k_2}) \in E_{TG'}$). Fig.\ref{fig:rcg-cycle-1} shows an illustration of this, where a dashed arrow represents an edge or a path and solid arrows represent edges.
While there is no overlap between the lifetime of $tl^{i_1}_{j_1}$ and $tl^{i_2}_{j_2}$, the hardware resources occupied by the time layer $tl^{i_2}_{j_2}$ must be reconfigured to be the next time layer in the same DRR, $tl^{i_2}_{j_2+1}$  before the configuration of $tl^{i_1}_{j_1}$, and there must be an edge from $m_{k_2}$ to $tl^{i_2}_{j_2+1}$ (shown in a bold dashed arrow in Fig.\ref{fig:rcg-cycle-1}) because a time layer can only be configured after the execution of all tasks in the earlier time layers in the same DRR. Accordingly, a cycle is formed, which indicates the conflict of constraints.

\textit{ONLY IF}. Here, we show that if the given partition and configuration order is infeasible, there must be backward dependencies between the time layers that have no lifetime overlap.

If the given partition and configuration order is infeasible, there must be a cycle in $RCG$. \revise{Notice} that $V_{RCG} = V_{TG}\cup V_{LTG}$.
The subgraph induced by $V_{LTG}$, which includes the edge set $E_{cr}$ representing the configuration order of the time layers, is acyclic.
The subgraph induced by $V_{TG}$ (exactly $TG$), which includes the edge set $E_{TG}$ representing the dependences between tasks, is also acyclic.
\revise{Moreover}, all the edges in $E_{ce}$ are from  $V_{LTG}$ to $V_{TG}$, which represents that a task must be configured before it is executed, and all the edges in $E_{ec}$ are from $V_{TG}$ to $V_{LTG}$, which represents that a time layer can only be configured after the execution of all the tasks in the earlier time layers. Consequently, the cycle must include four parts: \textbf{1)} a path (one or more edges) from $E_{cr}$, \textbf{2)} a path (one or more edges) from $E_{TG}$, \textbf{3)} an edge from $E_{ce}$, and \textbf{4)} an edge from $E_{ec}$.

Without loss of generality, we assume that the cycle includes a path from $tl^{i_1}_{j_1}$ to $tl^{i_2}_{j_2}$ and a path from $m_{k_1}$ to $m_{k_2}$, respectively, constructed by the edges from $E_{cr}$ and $E_{TG}$.
The cycle must also include two edges:  $(m_{k_2}, tl^{i_1}_{j_1})$ and $(tl^{i_2}_{j_2}, m_{k_1})$. Fig.\ref{fig:rcg-cycle-2} shows an illustration of this.
According to the definition of $E_{ce}$, $m_{k_1}$ is in $tl^{i_2}_{j_2}$ because we have the edge $(tl^{i_2}_{j_2}, m_{k_1})$.
On the other hand, the edge $(m_{k_2}, tl^{i_1}_{j_1})$ indicates that $tl^{i_1}_{j_1}$ is configured after $m_{k_2}$ is executed, which means that $m_{k_2}$ must be located in the previous time layer of $tl^{i_1}_{j_1}$ in $drr_{i_1}$, $tl^{i_1}_{j_1-1}$. We can see that $tl^{i_1}_{j_1-1}$ and $tl^{i_2}_{j_2}$ have a backward data dependence and their lifetimes are non-overlapping, as the region occupied by $tl^{i_1}_{j_1-1}$ has been reconfigured to be $tl^{i_1}_{j_1}$ before $tl^{i_2}_{j_2}$ is configured.

\hfill \textbf{Proof END}.

Note that if $RS$ represents a topological ordering of $TG$, the partition and configuration order will always be feasible because there are no backward dependencies involved.
\begin{corollary}
\label{the:liftimeoverlap}
Given a partition, a configuration order, and the task dependency graph, the $RCG$ is acyclic if there is always lifetime overlap between time layers that have backward dependencies.
\end{corollary}

\begin{figure}[htbp]
	\centering
    \subfloat[An example of RCG under feasible partition and configuration order.]{
        \includegraphics[width=0.48\textwidth, height=0.20\textwidth]{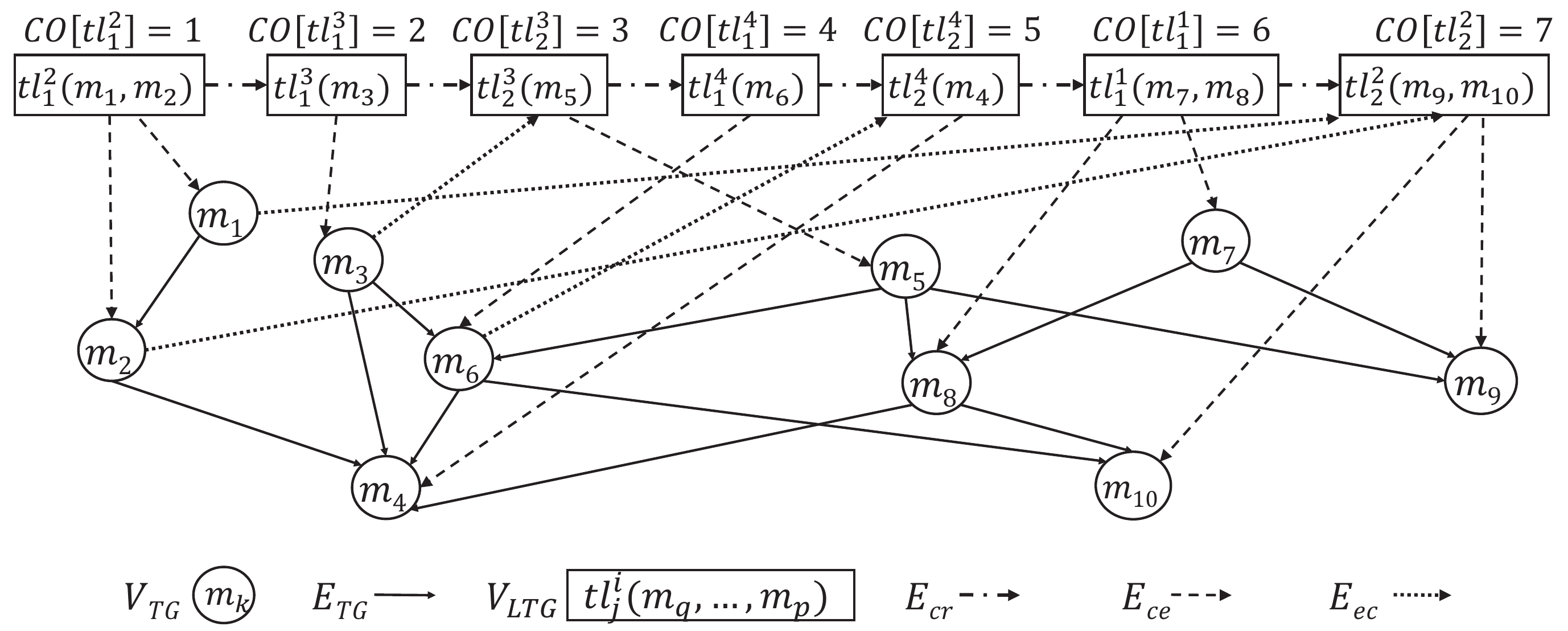}
		\label{fig:RCG2}
	}
    \vspace{0.5em}	
    \subfloat[An example of RCG with a cycle (No feasible schedule).]{
        \includegraphics[width=0.48\textwidth, height=0.20\textwidth]{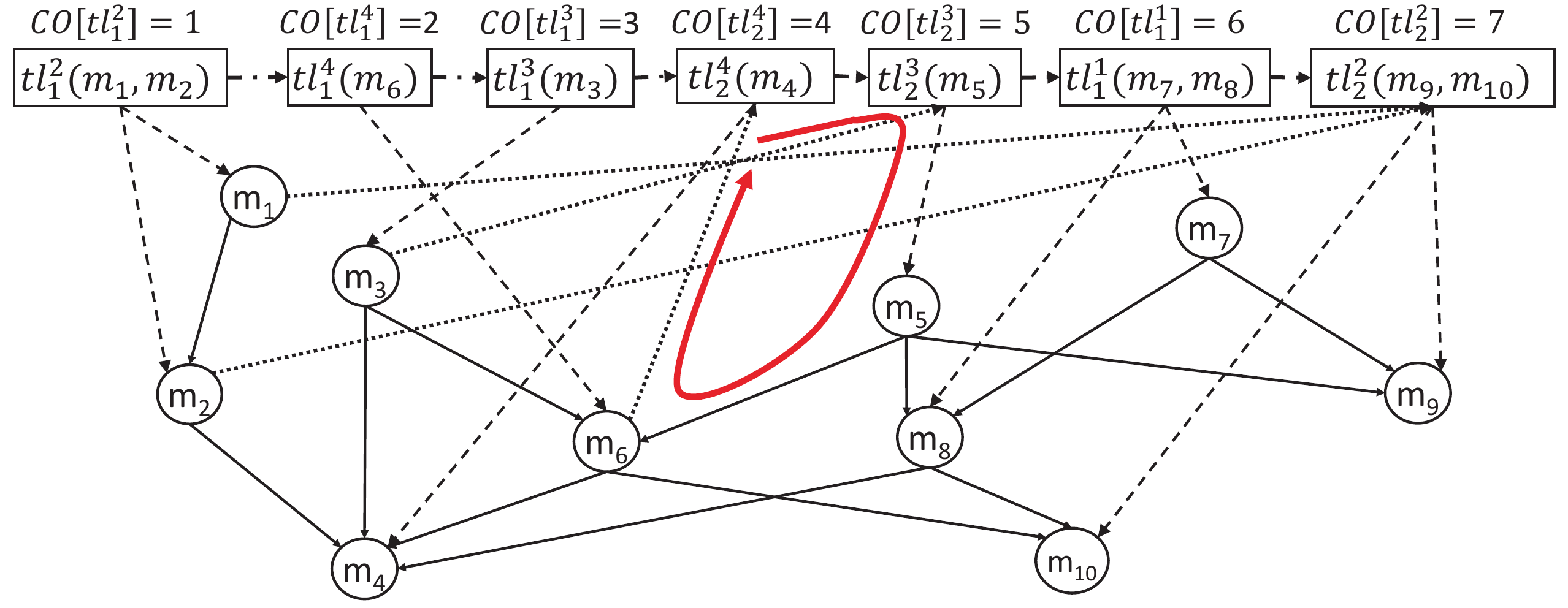}
		\label{fig:illegal}
	}
	\caption{RCG Examples}
	\label{fig:RCG-examples}
\end{figure}

Fig. \ref{fig:RCG2} shows the $RCG$ of the feasible $P$-$ST$ in Formula (\ref{equ:p-st-example}), where
$RS=\langle(1\ 2)_1^2\ (3)_1^3\ (5)_2^3\ (6)_1^4\ (4)_2^4\ (7\ 8)_1^1\ (9\ 10)_2^2\rangle$.
If $RS$ is changed to the configuration order in Fig. \ref{fig:feasibility.eps}, $\langle(1\ 2)_1^2\  (6)_1^4\ (3)_1^3\ (4)_2^4\ (5)_2^3\  (7\ 8)_1^1\ (9\ 10)_2^2\rangle$, the corresponding $RCG$ is shown in Fig. \ref{fig:illegal}, where a cycle $tl^4_2\rightarrow tl^3_2\rightarrow m_5\rightarrow m_6\rightarrow tl^4_2$ is formed and no feasible schedule can be found.

\vspace{-1em}
\subsection{Computation of the Schedule}
\label{sec:rcg}
The schedule can be computed by finding the longest paths on the RCG with edges weighted as follows.
\begin{equation}
\label{equ:RCGwt}
  \forall tl^i_j\in V_{LTG}, \revise{wt}(tl^i_j) = c_{tl^i_j}, \forall m_i\in V_{TG}, \revise{wt}(m_i) = t_i.
\end{equation}

Let $s$ be the vertex corresponding to the time layer that is configured first (having zero in-degree in $RCG$),
and $lp(v_i)$ denote the vertex-weighted longest-path from $s$ to a vertex $v_i$.
\revise{The schedule ($bc_{tl^i_j}$ and $bt_{k}$)} can be determined by computing  $lp(tl^i_j)$ and $lp(m_k)$, respectively.

The \textit{schedule length} $T$ of the partially dynamically reconfigurable system is the maximum of the paths, and can be calculated as follows:
$T = \max\limits_{1 \leq k \leq n}(lp(m_k) + t_k).$

Given a feasible $P$-$ST$, we can construct RCG and compute the schedule in $O(n^2)$ time if the RCG is acyclic, where $n$ is the number of task modules.

\section{Optimization Framework}\label{sec:perturbation}
\begin{definition}
An \emph{insertion point} in the partitioned sequence triple $P$-$ST$$(PS, QS, RS)$ is defined as a four-tuple, $(p, q, r, tl^i_j)$, where $p$, $q$, and $r$ are the positions immediately after the $p$-$th$ task module in $PS$, the $q$-$th$ task module in $QS$, and the $r$-$th$ task module in $RS$, respectively, and $tl^i_j$ is the $j$-th time layer in $drr_i$. $p$ = 0 (or $q$ = 0 or $r$ = 0) indicates the position before the first task module of the sequence.
\end{definition}

In Section \ref{subsec:fip}, we will discuss the feasibility and types of \textit{insertion points} in detail.

\vspace{-1em}
\subsection{Overall Design Flow}\label{sec:overall}
In this work, we modify the perturbation method, Insertion-after-Remove (IAR) in \cite{chen2008fixed}, to explore the design space of the schedule and floorplan in a simulated annealing-based search. With the IAR operation, we can perturb the partitioning, scheduling, and floorplanning of task modules simultaneously. 
The detailed steps are as follows:
\begin{enumerate}
\item[a.] Select and remove a task module $m_{k}$ randomly and then compute the floorplan and schedule of task modules without the removed task module $m_{k}$;
\item[b.] Select a fixed number of feasible candidate \emph{insertion points}, $S_{CIP}= \{(p, q, r, tl^i_j)\}$, for $m_{k}$ by rough evaluations of all the feasible \emph{insertion points};
\item[c.] Choose the best \emph{insertion point} from $S_{CIP}$ for the removed task module $m_{k}$ by accurate evaluations.
\end{enumerate}

In step $b$, the feasible \emph{insertion points} are evaluated by the linear combination of resource costs, \textit{schedule length}, and communication costs.
In this step, the resource costs are calculated accurately.
To reduce the time complexity, the communication cost $CC$ is calculated roughly without updating the floorplan and schedule of task modules, and the schedule length is roughly evaluated by under-estimating the configuration spans of time layers using Formula \ref{equ:rough_ct}. In step $c$, all the \emph{insertion points} in $S_{CIP}$ will be evaluated accurately based on the entire floorplan \revise{considering the communication costs}, and the best one will be chosen as the candidate \emph{insertion point}. The feasibility of \emph{insertion points} will be discussed in Subsection \ref{sec:feas-st}.

In the experiments, we set the size of $|S_{CIP}|$ at 15.
The objective function $Cost$ is defined as the linear combination of the area cost ($AC$), which depends on the dimensions of all occupied resources ($Col\times Row$), the \textit{schedule length} ($T$), and the communication costs ($CC$):
\begin{equation}\label{equ:evl}
Cost=\alpha\times AC + \beta\times T + \gamma\times CC.
\end{equation}

\revise{$\alpha$, $\beta$, and $\gamma$ are balance factors for making a trade-off between the schedule length and the communication costs. T is calculated by under-estimating the configuration spans of time layers using Formula (\ref{equ:rough_ct}).}
The evaluation of $T$, $CC$, and $AC$ will be discussed in Subsection \ref{sec:evaluation}.

\subsection{Feasible insertion points in $P$-$ST$}
\label{sec:feas-st}
Generally, given a $P$-$ST$ of $n-1$ task modules, there are a total of $O(n^{3})$ \emph{insertion points} for inserting a task module. However, when considering Theorem~\ref{the:feasibility} and the definition of $P$-$ST$, some \emph{insertion points} are infeasible.
Here we discuss the feasibility of \emph{insertion points} in $P$-$ST$s.

\subsubsection{Lifetime overlap constraint}\label{subsec:life}

First, inserting $m_{k}$ could introduces new backward dependencies between time layers. To ensure the lifetime overlap between the backward-dependent time layers, we have the following corollary from Theorem \ref{the:feasibility}.

\begin{corollary}
\label{cor:lifetime-mk}
The lifetime of a time layer, $tl^{i}_{j}$, where $m_{k}$ is inserted, must satisfy the following condition.
\begin{equation}
\label{equ:lifetime-mk}
\begin{cases}
	lt\_s^{i}_{j} \le 
	\min_{tl^{i_1}_{j_1}:\exists m_{k_1}\in tl^{i_1}_{j_1}\wedge(m_{k},m_{k_1})\in E_{TG'} }\{lt\_e^{i_1}_{j_1}\};\\
	lt\_e^{i}_{j} \ge 
	\max_{tl^{i_1}_{j_1}:\exists m_{k_1}\in tl^{i_1}_{j_1} \wedge(m_{k_1},m_{k})\in E_{TG'} }\{lt\_s^{i_1}_{j_1}\}.
	\end{cases}
\end{equation}
\end{corollary}

Second, the lifetime of a time layer is changed when a new time layer is inserted into an existing DRR.
To ensure the lifetime overlap between the time layers that have backward dependences, we have the following corollary from Theorem \ref{the:feasibility}.

\begin{corollary}
\label{cor:lifetime-min}
	Given a partition and configuration order, the lifetime of a time layer $tl_j^i$ must satisfy the following \textit{minimum lifetime constraint}, denoted as $LT^{min}[tl^i_j]=(lt\_ls^i_j,lt\_ee^i_j)$.

$lt\_ls^i_j = lt\_s^i_j$.
\begin{equation}
    lt\_ee^i_j =
    \begin{cases}
     \max_{tl^{i'}_{j'}:(tl^{i'}_{j'}, tl^i_j)\in E_{LTG}\wedge lt\_s^{i'}_{j'}>lt\_s^{i}_{j}}\{lt\_s^{i'}_{j'}\} + 1,\\
     ~~\text{if there is a backward dependence} \\
     ~~\text{between $tl_j^i$ and some future time layer.}\\
     lt\_s^i_j + 1, \text{otherwise.}
     \end{cases}
\end{equation}
\end{corollary}

The minimum lifetime constraints ensure a lifetime overlap between any two time layers that have a backward dependence.
This constraint cannot be violated after $m_{k}$ is inserted back into the $P$-$ST$. In Section \ref{subsec:example}, an example is provided.

\subsubsection{Feasibility of insertion points}
\label{subsec:fip}

Let $ps[i]$, $qs[i]$, and $rs[i]$, $1 \le i < n$ represent the $i$-$th$ task in $PS$, $QS$, and $RS$, respectively, with $m_k$ removed.
For each possible \emph{insertion point} $(p, q, r, tl^i_j)$, there exist three possible types of optional partitions to re-insert $m_k$ depending on the time layer $tl^i_j$.

\textit{Type-1}:
Create a new time layer in a new DRR, $tl^{new}_{new}$, for $m_k$. In this case, $(p, q, r)$ must be located within the boundary of task sequences corresponding to different DRRs in $P$-$ST$, i.e.,
\begin{equation}
\label{equ:ip-type1}
\begin{split}
&drr(ps[p])\!\ne\! drr(ps[p+1]),~drr(qs[q])\!\ne\! drr(qs[q+1]), \\
&and~tl(rs[r])\!\ne\! tl(rs[r+1]).
\end{split}
\end{equation}

Without loss of generality, we assume that $ps[0] = ps[n] = -1$, and that $drr(-1)$ and $tl(-1)$ correspond to a virtual DRR and to virtual time layers, respectively. $qs[0]$, $qs[n]$, $rs[0]$, and $rs[n]$ are dealt with similarly.

This type of insertion point will not change the lifetimes of any other time layers according to Definition \ref{equ:lifetime-def}.
Consequently, if the constraint (\ref{equ:lifetime-mk}) in Corollary \ref{cor:lifetime-mk} is satisfied, then $(p, q, r, tl^{new}_{new})$ is feasible. Note that the new generated time layer $tl^{new}_{new}$ is configured between $tl(rs[r])$ and $tl(rs[r+1])$.

\textit{Type-2}: Create a new time layer, $tl^{i}_{new}$, in an existing DRR, $drr_i$, for  $m_k$. In this case, the insertion point $(p, q, r, tl^{i}_{new})$ must be located within the boundary of task sequences corresponding to different time layers, i.e.,
there is a combination $(p',q',r')\in \{p, p+1\}\times\{q, q+1\}\times \{r, r+1\}$ such that $drr(ps[p'])\!=\! drr(qs[q'])\!=\! drr(rs[r'])\!\ne\!drr(-1)$, and
\[
\begin{split}
&tl(ps[p])\!\ne\! tl(ps[p+1]),~tl(qs[q])\!\ne\! tl(qs[q+1]),\\
&and~tl(rs[r])\!\ne\! tl(rs[r+1]).
\end{split}
\]

This type of insertion point will change the lifetime of the time layer that is immediately before $tl^{i}_{new}$ in $drr_i$. An insertion point $(p, q, r, tl^{i}_{new})$ is feasible if the constraints in both Corollary \ref{cor:lifetime-mk} and Corollary \ref{cor:lifetime-min} are satisfied. Note that the new generated time layer $tl^i_{new}$ is configured between $tl(rs[r])$ and $tl(rs[r+1])$.

\textit{Type-3}: Insert $m_k$ into an existing time layer, $tl^i_j$. In this case, an insertion point $(p, q, r, tl^i_j)$ must satisfy the condition that there is a combination $(p',q',r')\in \{p, p+1\}\times\{q, q+1\}\times \{r, r+1\}$, such that
$tl(ps[p'])= tl(qs[q'])=tl(rs[r'])\!\ne\!tl(-1)$.

This type of insertion point will not change the lifetime of any other time layers.  $(p, q, r, tl^i_j)$ is feasible if the constraint (\ref{equ:lifetime-mk}) in Corollary \ref{cor:lifetime-mk} is satisfied.

\subsubsection{An example}\label{subsec:example}
Given the task dependencies shown in Fig.\ref{fig:TG}, we have the following $P$-$ST$ with $m_8$ removed:
\begin{equation}
\label{equ:no-mk}
\begin{split}
&(\langle[(1\ 2)_1^2\ (9\ 10)_2^2]_2\ [(7)_1^1]_1\ [(6)_1^4\ (4)_2^4]_4\ [(3)_1^3\ (5)_2^3]_3\rangle,\\
&\langle[(7)_1^1]_1\ [(2\ 1)_1^2\ (9\ 10)_2^2]_2\ [(3)_1^3\ (5)_2^3]_3\ [(6)_1^4\ (4)_2^4]_4\rangle,\\
&\langle(1\ 2)_1^2\ (3)_1^3\ (5)_2^3\ (6)_1^4\ (4)_2^4\ (7)_1^1\ (9\ 10)_2^2\rangle).
\end{split}
\end{equation}

Fig. \ref{fig:insertpoint} (black edges) shows the lifetime of time layers with the task module $m_8$ removed.

According to Corollary \ref{cor:lifetime-mk}, for the lifetime, ($lt\_s^{i}_{j}, lt\_e^{i}_{j})$, of a time layer $tl^{i}_{j}$, where the removed task module $m_8$ will be inserted, we have the following basic constraints.
\begin{enumerate}
\item[(i)] Because ($m_8$, $m_4$) and ($m_8$, $m_{10}$) are in $E_{TG'}$, and $m_4$ is in  $tl^4_2$ and $m_{10}$ is in $tl^2_2$, then $lt\_s^{i}_{j}$ has to satisfy:

$lt\_s^{i}_{j} \le \min\{lt\_e^{4}_{2}, \ lt\_e^{2}_{2}\}$.

As shown in Fig. \ref{fig:insertpoint}, $LT[tl^4_2]=(5,\infty)$ and $LT[tl^2_2]=(7,\infty)$, thus, $lt\_s^{i}_{j} \le \min\{\infty,\ \infty\} = \infty$, which means that there is no constraint on the beginning of the lifetime.

\item[(ii)] Because ($m_5$, $m_8$) and ($m_7$, $m_8$) are in $E_{TG'}$, and $m_5$ is in  $tl^3_2$ and $m_{7}$ is in $tl^1_1$, then $lt\_e^{i}_{j}$ has to satisfy:

$lt\_e^{i}_{j} \ge \max\{lt\_s^{3}_{2}, \ lt\_s^{1}_{1}\}$.

As shown in Fig. \ref{fig:insertpoint}, $LT[tl^3_2]=(3,\infty)$ and $LT[tl^1_1]=(6,\infty)$, thus, $lt\_e^{i}_{j} \ge \max\{3,\ 6\} = 6$.
\end{enumerate}

Consequently, the task module $m_8$ must be inserted into a time layer whose lifetime ends after $6$.

Here, we assume that $m_8$ is inserted back into an \emph{insertion point}, $(6^{th}, 8^{th}, 5^{th}, tl^i_j)$, of the $P$-$ST$$(PS, QS, RS)$ shown in Formula \ref{equ:no-mk}.
For the \emph{insertion point} $(6^{th}, 8^{th}, 5^{th}, tl^i_j)$, there are several optional partitions for $m_8$ considering different $tl^i_j$.

For a \textit{Type-1} partition because the task modules $ps[6]=m_6$ and $ps[7]=m_4$ belong to the same DRR $drr_4$, i.e., $drr[m_6] = drr[m_4]$, we cannot create a new DRR for $m_8$.

For a \textit{Type-2} partition, a new time layer $tl_{new}^{4}$ is created for $m_8$ in the existing DRR $drr_4$ as follows.
\begin{equation*}
\begin{split}
&(\langle[(1\ 2)_1^2\ (9\ 10)_2^2]_2\ [(7)_1^1]_1\ [(6)_1^4\ (\textbf{8})_{new}^{4}\ (4)_2^4]_4\ [(3)_1^3\ (5)_2^3]_3\rangle,\\
&\langle[(7)_1^1]_1\ [(2\ 1)_1^2\ (9\ 10)_2^2]_2\ [(3)_1^3\ (5)_2^3]_3\ [(6)_1^4\ (\textbf{8})_{new}^{4}\ (4)_2^4]_4\rangle,\\
&\langle(1\ 2)_1^2\ (3)_1^3\ (5)_2^3\ (6)_1^4\ (\textbf{8})_{new}^{4}\ (4)_2^4\ (7)_1^1\ (9\ 10)_2^2\rangle).
\end{split}
\end{equation*}

Without loss of generality, we set the configuration order $CO[tl^4_{new}]$ of time layer $tl_{new}^{4}$ at 4.5 in this situation,
and the lifetime of $tl_{1}^{4}$ will be changed to (4, 4.5).
Fig. \ref{fig:insertpoint} (black edges and red edges) shows the lifetime of the time layers. As the lifetime of the inserted time layer $tl_{new}^{4}$ is $LT[tl^4_{new}]=(4.5, 5)$, where the end of the lifetime is $lt\_e^{4}_{new} = 5 < 6$, the insertion point $(6^{th}, 8^{th}, 5^{th}, tl_{new}^{4})$ is infeasible.

For a \textit{Type-3} partition, the removed task module $m_8$ is inserted into the existing time layer $tl_{1}^{4}$ or $tl_{2}^{4}$.

For $tl^i_j = tl_{1}^{4}$, the $P$-$ST$ is as follows.
\begin{equation*}
\begin{split}
&(\langle[(1\ 2)_1^2\ (9\ 10)_2^2]_2\ [(7)_1^1]_1\ [(6\ {\textbf{8}})_1^4\ (4)_2^4]_4\ [(3)_1^3\ (5)_2^3]_3\rangle,\\
&\langle[(7)_1^1]_1\ [(2\ 1)_1^2\ (9\ 10)_2^2]_2\ [(3)_1^3\ (5)_2^3]_3\ [(6\ {\textbf{8}})_1^4\ (4)_2^4]_4\rangle,\\
&\langle(1\ 2)_1^2\ (3)_1^3\ (5)_2^3\ (6\ {\textbf{8}})_1^4\ (4)_2^4\ (7)_1^1\ (9\ 10)_2^2\rangle).
\end{split}
\end{equation*}

In this situation, however, as shown in Fig. \ref{fig:insertpoint}, the lifetime of the inserted time layer $tl_{1}^{4}$ is $LT[tl^4_1]=(4,5)$, where the end of the lifetime is $lt\_e^{4}_{1} = 5 < 6$, thus, the insertion point $(6^{th}, 8^{th}, 5^{th}, tl_{1}^{4})$ is infeasible.

For $tl^i_j = tl_{2}^{4}$, the $P$-$ST$ is as follows.
\begin{equation*}
\begin{split}
&(\langle[(1\ 2)_1^2\ (9\ 10)_2^2]_2\ [(7)_1^1]_1\ [(6)_1^4\ ({\textbf{8}}\ 4)_2^4]_4\ [(3)_1^3\ (5)_2^3]_3\rangle,\\
&\langle[(7)_1^1]_1\ [(2\ 1)_1^2\ (9\ 10)_2^2]_2\ [(3)_1^3\ (5)_2^3]_3\ [(6)_1^4\ ({\textbf{8}}\ 4)_2^4]_4\rangle,\\
&\langle(1\ 2)_1^2\ (3)_1^3\ (5)_2^3\ (6)_1^4\ ( {\textbf{8}}\ 4)_2^4\ (7)_1^1\ (9\ 10)_2^2\rangle).
\end{split}
\end{equation*}

In this situation, as shown in Fig. \ref{fig:insertpoint}, the lifetime of the inserted time layer $tl_{2}^{4}$ is $LT[tl^4_2]=(5,\infty)$, where the end of the lifetime is $lt\_e^{4}_{2} = \infty \ge 6$, thus, the insertion point $(6^{th}, 8^{th}, 5^{th}, tl_{2}^{4})$ is feasible.
\begin{figure}[htbp]
    \centering
    \includegraphics[width=0.50\textwidth,height=0.30\textwidth]{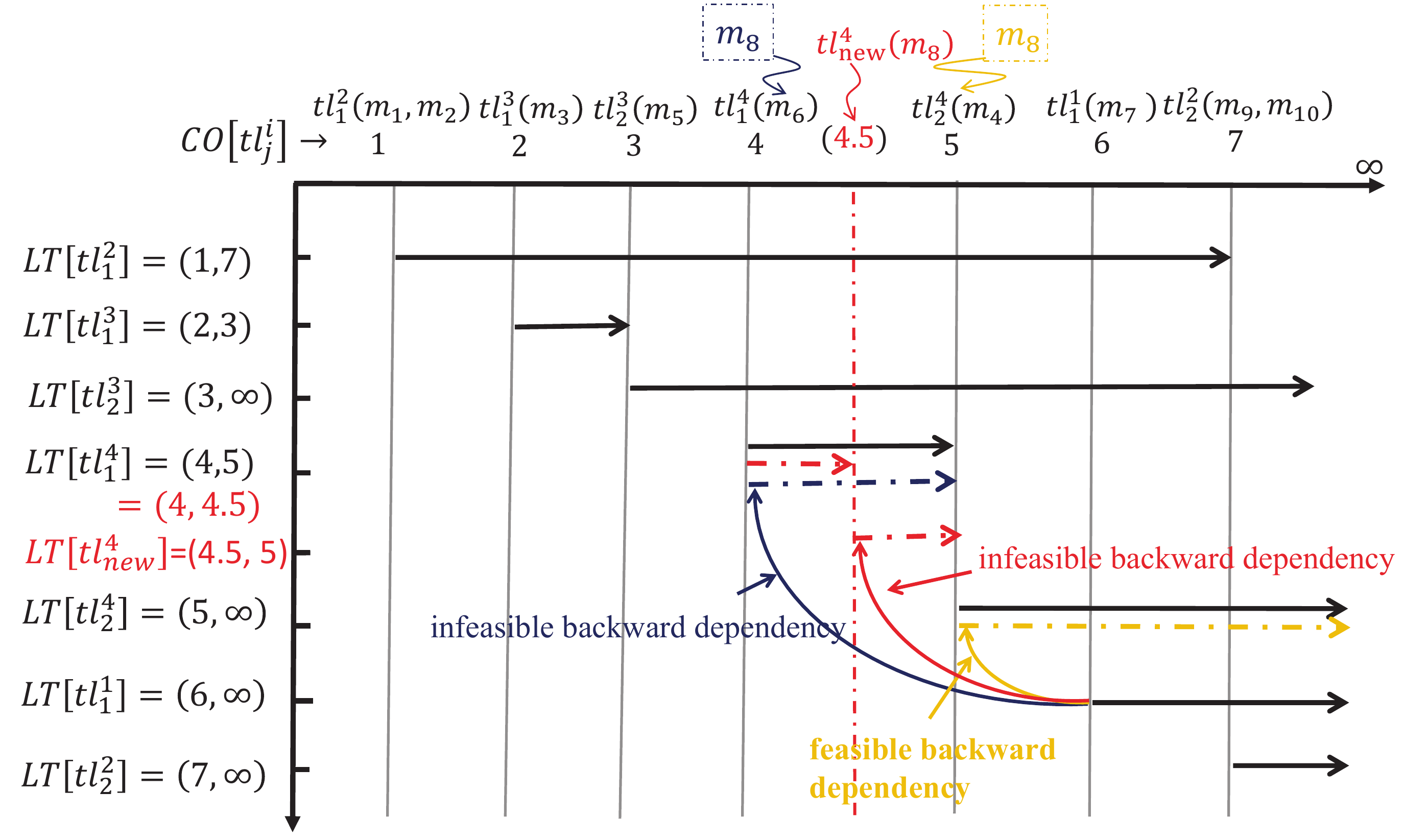}
    \caption{ Example of feasible positions for inserting $m_8$.}
    \label{fig:insertpoint}
\end{figure}

\subsubsection{Discussion of the Reachability of the Solution Space }
\label{subsec:reachability}
\begin{theorem} Any two feasible solutions of $n$ task modules, represented by $P$-$ST$s, are reachable to each other through at most \revise{$2n$} feasible solutions generated by iteratively removing and re-inserting a task module.
\end{theorem}
\textbf{Proof}. As discussed in Section \ref{subsec:sn-condition}, if the RS is a topological order of the task dependence graph, the $P$-$ST$ will be feasible.
From any feasible solution $P'$-$ST'$: $(PS', QS', RS')$, we are able to reach another feasible solution $P$-$ST$: $(PS, QS, RS)$, where $RS$ is a topological order, by iteratively removing and re-inserting some task module.
If we select the task modules for removing and inserting back in the order of $RS'$, no backward dependencies between time layers will be introduced because all the time-layer dependencies introduced by the moved task module are forward ones (per Definition \ref{def:forward}).
Consequently, all the intermediate solutions generated from $P$-$ST$, assumed to be $P$-$ST_1$, $P$-$ST_2$, $\cdots$, $P$-$ST_k$ ($k\le n$), will always be feasible according to Theorem \ref{the:feasibility}\revise{, and from one solution $P$-$ST'$ ($PS'$, $QS'$, $RS'$) we can reach any other solution $P$-$ST$ ($PS$, $QS$, $RS$) where $RS$ is a topological order of $TG$ in at most $k$ ($k\le n$) steps: $P$-$ST'$, $P$-$ST_1$, $P$-$ST_2$ $\cdots$ $P$-$ST_k$ = $P$-$ST$.}

\revise{On the other hand, we can reach a generic feasible solution $P$-$ST''$ ($PS''$, $QS''$, $RS''$) with no constraints on $RS''$ from any solution $P$-$ST$ ($PS$, $QS$, $RS$) where $RS$ is a topological order through at most $n$ feasible solutions, which can be obtained by inverting the sequence of removing and re-inserting operations from $P$-$ST$ to $P$-$ST''$.}

Therefore, \revise{starting from one generic solution $P$-$ST'$ we can reach another generic solution $P$-$S''$ (through a solution $P$-$ST$ in which $RS$ is a topological order of TG) in at most $2n$ steps.}

\revise{\hfill \textbf{Proof END}.}

\vspace{-0.5em}
\subsection{Evaluation of Insertion Points}
\label{sec:evaluation}

\subsubsection{Computation of ~$T$}
To reduce the complexity, we use the area sum of the task modules to underestimate the configuration span of the time layers instead of accurately computing the configuration span of the DRRs. In this subsection, we discuss a method (used in step $b$ of the IAR perturbation shown in Section \ref{sec:perturbation}) to evaluate, in amortized constant time, the \textit{schedule length} while inserting a task module into an insertion point.

After a task module $m_k$ is removed from the partitioned sequence triple $P$-$ST$, $RCG$ is updated by removing some edges related to time layers according to the following two situations:
\begin{enumerate}
\item[(i)] If $m_k$ is the only task module in $tl(m_k)$, we remove the vertex $tl(m_k)$ along with its incoming and outgoing edges and the edges between $m_k$ and any vertices that represent time layers.
\item[(ii)] If $m_k$ is not the only task module in time layer $tl(m_k)$, the weight of vertex $tl(m_k)$ will be subtracted by the configuration time span of \textcolor{blue}{task} $m_k$ ($c_{k}$) and the edges between $m_k$ and any vertices that represent time layers are removed.
\end{enumerate}

Let $RCG^0$ be the updated reconfiguration constraint graph. To simplify the description, we add to $RCG^0$ a source vertex $v_s$ with outgoing edges to all the task modules that have zero in-degree and a sink vertex $v_t$ with incoming edges from all the task modules that have zero out-degree. Both $v_s$ and $v_t$ have zero weight. Let $rRCG^0$ be the graph obtained by reversing all the edges of $RCG^0$.

We pre-compute the longest paths from $v_s$ to each vertex $v_i\in RCG^{0}$, denoted as $lp_{0}(v_i)$, and the longest paths from $v_t$ to each vertex, $lp^r_0(v_i)$, based on $rRCG^0$ in $O(n^2)$ time using the longest-path algorithm on directed acyclic graphs.

To evaluate the \textit{schedule length} $T$ for inserting $m_k$ into a feasible \emph{insertion point} $(p,q,r, tl^i_j)$ in $P$-$ST$,
a new reconfiguration constraint graph $RCG^{new}$ is generated.
Let $T_0$ be the longest path from $v_s$ to $v_t$ in $RCG^{0}$.
$T$ can be roughly and incrementally evaluated from the longest paths in $RCG^{0}$ and $rRCG^0$ by considering only the paths passing through the vertex $tl^i_j$ or $m_k$ because all the changed edges are related to either $tl^i_j$ or $m_k$.
\begin{equation}
\label{eq:STime}
\begin{split}
T =\max (T_0,~&lp_{new}(tl^i_j)+lp^r_{new}(tl^i_j)+c_{tl^i_j},\\
~&lp_{new}(m_k)+lp^r_{new}(m_k)+t_k),
\end{split}
\end{equation}
where $lp_{new}(tl^i_j)$, $lp^r_{new}(tl^i_j)$, $lp_{new}(m_k)$, and $lp^r_{new}(m_k)$ are incrementally computed based on $lp_{0}(v_i)$ and $lp^r_0(v_i)$, for the three types of partitions discussed in Section \ref{sec:feas-st}.

\begin{figure*}[htbp]
\centering
    \subfloat[Type-1: $tl^i_j$ is a new time layer in a new DRR.]{
        \includegraphics[width=0.24\textwidth, height=0.16\textwidth]{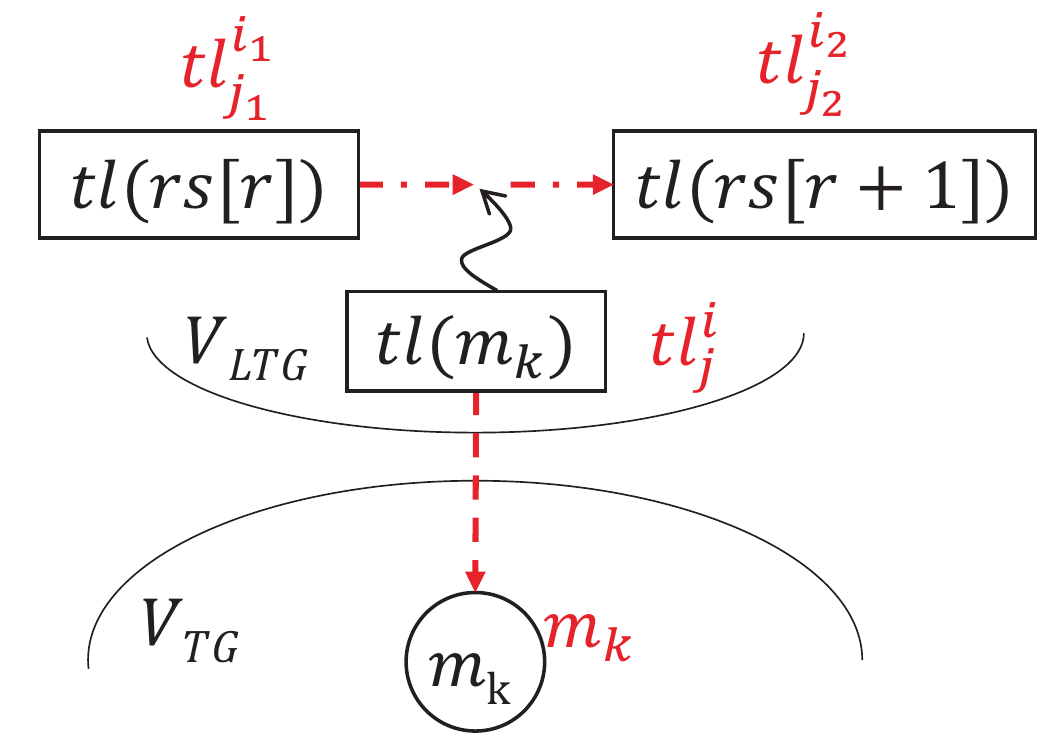}
        \label{fig:stime-type1}
    }
    \hspace{0.2em}
    \subfloat[Type-2: $tl^i_j$ is the first time layer in $drr_i$]{
        \includegraphics[width=0.28\textwidth,height=0.15\textwidth]{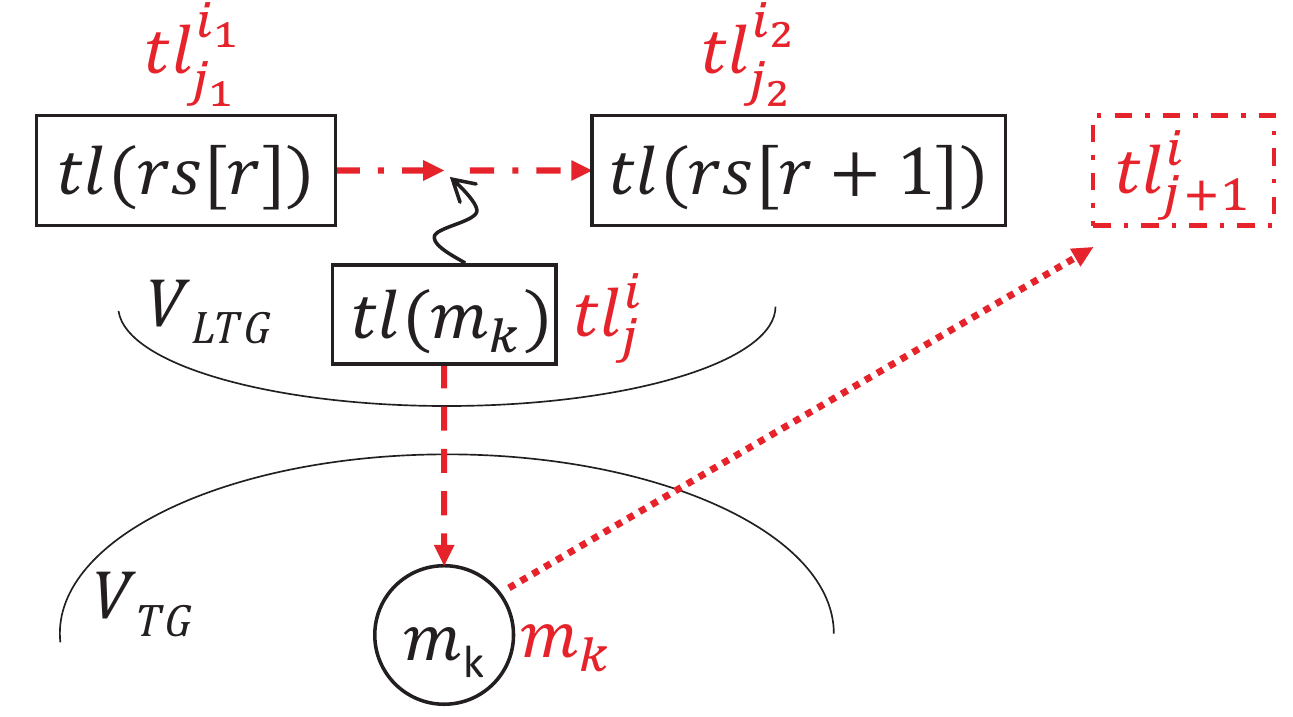}
        \label{fig:stime2a}
    }
    \hspace{0.2em}
    \subfloat[Type-2: $tl^i_j$ is the last time layer in $drr_i$]{
        \includegraphics[width=0.25\textwidth,height=0.16\textwidth]{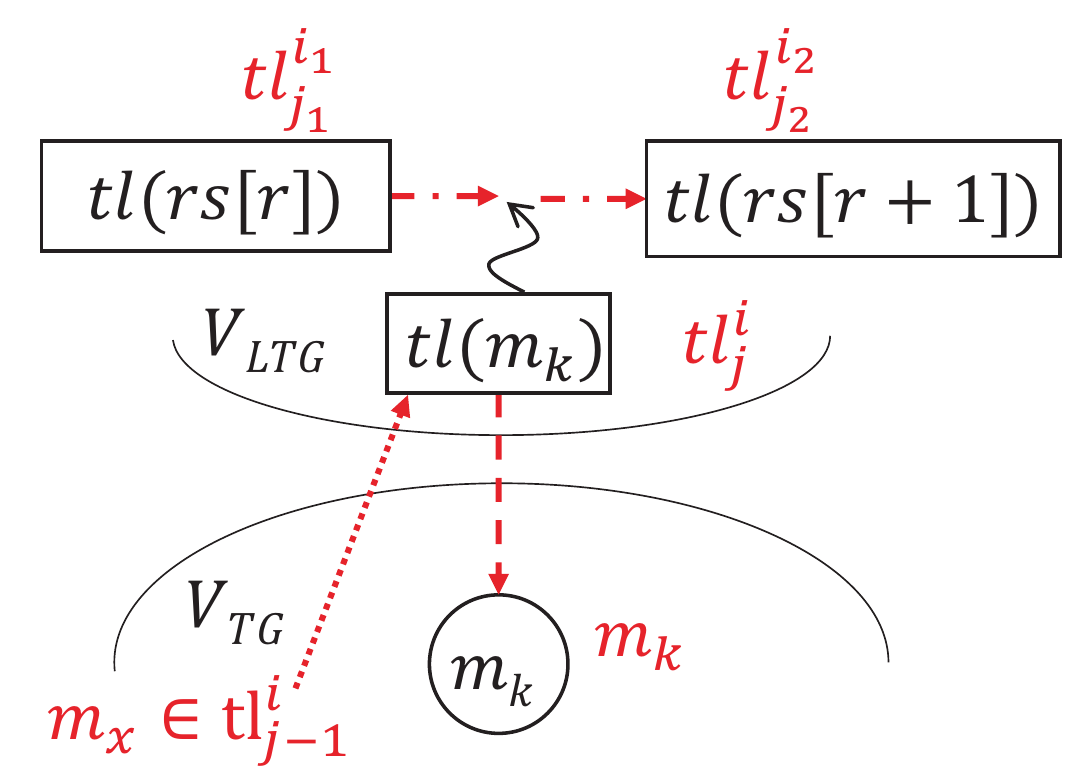}
        \label{fig:stime2b}
    }
    \vspace{0.2em}
    \subfloat[Type-2: $tl^i_j$ is the middle time layer in $drr_i$]{
        \includegraphics[width=0.32\textwidth,height=0.16\textwidth]{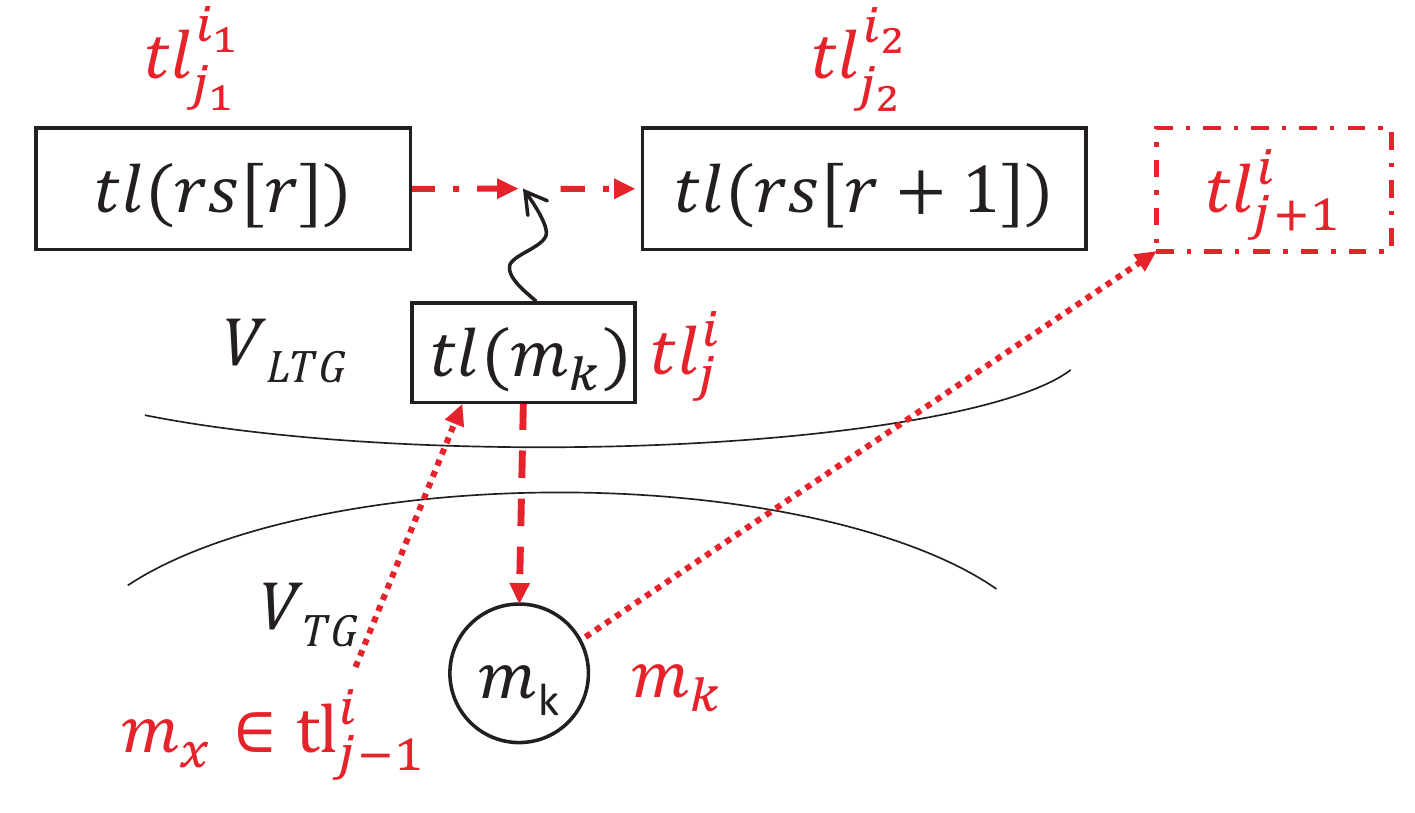}
        \label{fig:stime2c}
    }
    \hspace{0.3em}
    \subfloat[Type-3: $tl^i_j$ is the last time layer in $drr_i$]{
        \includegraphics[width=0.24\textwidth,height=0.15\textwidth]{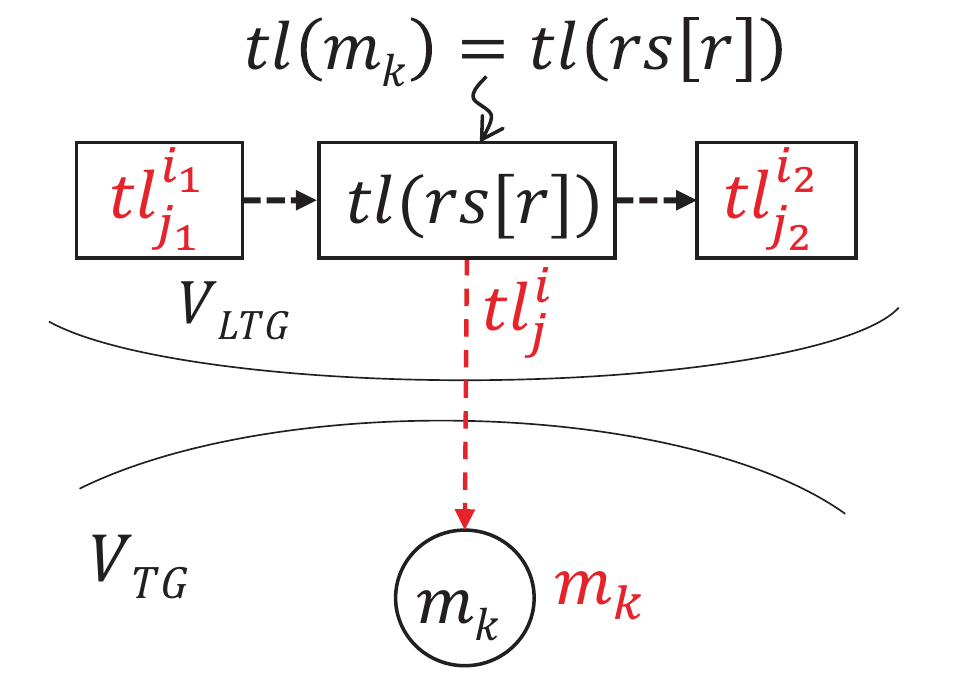}
        \label{fig:stime3a}
    }
    \hspace{0.3em}
    \subfloat[Type-3: $tl^i_j$ is not the last time layer in $drr_i$]{
        \includegraphics[width=0.27\textwidth,height=0.15\textwidth]{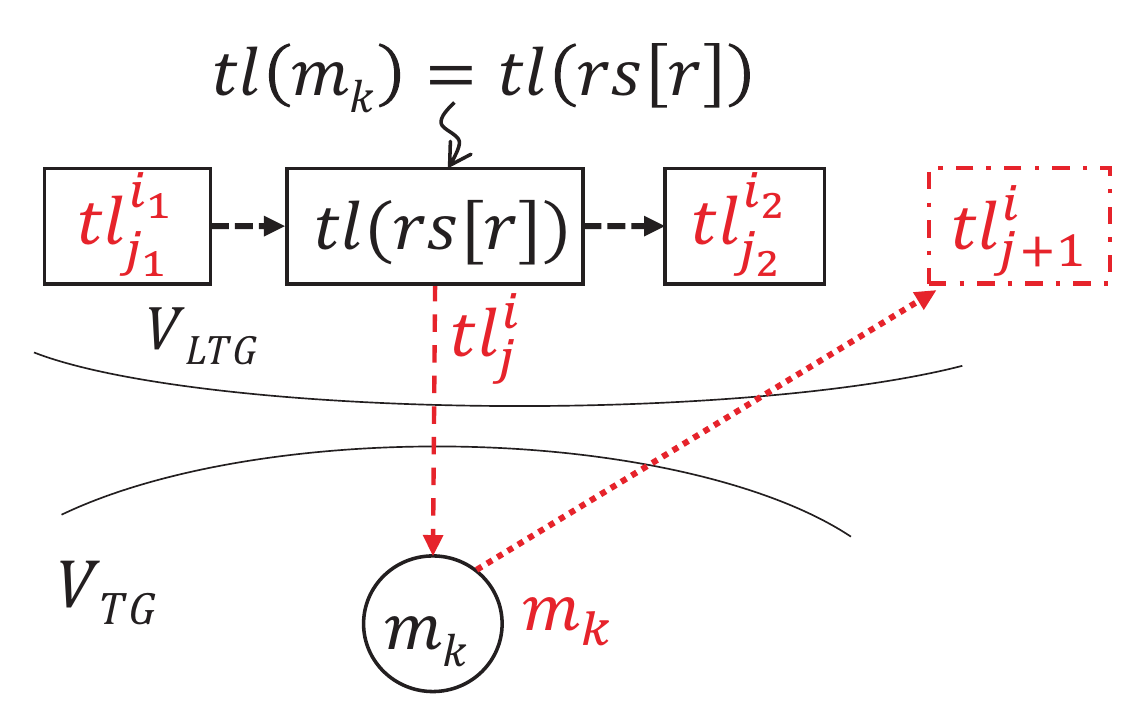}
        \label{fig:stime3b}
    }
\caption{$RCG^{new}$ for three types of partitions considering \textit{insertion points}.}
\label{fig:stime-type2}
\end{figure*}

\textit{Type-1}: Both a new DRR $drr_i$ and a new time layer $tl^i_j$ are created for $m_k$, and the $RCG^{new}$ can be constructed by adding three edges (red dotted lines) in $RCG^{0}$, as shown in Fig. \ref{fig:stime-type1}.

\textit{Type-2}: A new time layer $tl^i_j$ is created in an existing DRR $drr_i$ for $m_k$, and the $RCG^{new}$ can be constructed by adding some edges (red dotted lines) in $RCG^{0}$, where there are three situations respectively shown in Fig. \ref{fig:stime2a}, Fig. \ref{fig:stime2b} and Fig. \ref{fig:stime2c}.
In the situation shown in Fig. \ref{fig:stime2c}, there are at least three time layers $tl^i_{j-1}$, $tl^i_j$, and $tl^i_{j+1}$ in DRR $drr_i$.

\textit{Type-3}: $m_k$ is inserted into an existing time layer $tl^i_j$ in the DRR $drr_i$. There are two situations for updating $RCG^{new}$, respectively shown in Fig. \ref{fig:stime3a} and Fig. \ref{fig:stime3b}.

In the $RCG^{new}$, the computations of $lp_{new}(tl^i_j)$ and $lp_{new}(m_k)$ are summarized as follows.
\begin{small}
\begin{equation}
\label{equ:lp-new-tl}
lp_{new}(tl^i_j) =
\begin{cases}
lp_0(tl^{i_1}_{j_1})+c_{tl^{i_1}_{j_1}}, ~for~(a), ~(b), ~(e), ~and~(f);\\
\max\{lp_0(tl^{i_1}_{j_1})+c_{tl^{i_1}_{j_1}},\max\limits_{m_x\in tl^i_{j-1}}\{lp_0(m_x)+ t_x\}\}, \\
~~~~~~~~~~~~~~~~~~~~~for~(c) ~and~ (d).
\end{cases}
\end{equation}
\begin{equation}
\label{equ:lp-new-mk}
lp_{new}(m_k) =
\begin{cases}
\max\{lp_0(m_k), lp_{new}(tl^i_j)+c_k\},\\
~~~~~~~~~~~for~(a), ~(b), ~(c), ~and~(d);\\
\max\{lp_0(m_k), lp_{new}(tl^i_j)+(c_{tl^i_j}+c_k)\}, \\
~~~~~~~~~~~~~~~for~\textcolor{blue}{(e) ~and~ (f)}.\\
\end{cases}
\end{equation}
\end{small}

In the second part of Formula (\ref{equ:lp-new-tl}), the computation can be performed in amortized constant time because the total number of $m_x$ is at most $n-1$.

The longest paths $lp_{new}^r(tl^i_j)$ and $lp_{new}^r(m_k)$ in the reverse graph of $RCG^{new}$, $rRCG^{new}$,  can be calculated in constant time as follows.
\begin{equation}
\label{equ:lp-r-new-mk}
lp_{new}^r(m_k) =
\begin{cases}
lp_0^r(m_k), ~for~(a), ~(c),  ~and~(e);\\
\max\{lp_0^r(m_k), lp_0^r(tl^{i}_{j+1})+c_{tl^{i}_{j+1}}\}, \\
~~~~~~~~~~~~~for~(b),~(d),~and~(f).
\end{cases}
\end{equation}
\begin{equation}
\label{equ:lp-r-new-tl}
lp_{new}^r(tl^i_j) =
\max\{lp_0^r(tl^{i_2}_{j_2})+c_{tl^{i_2}_{j_2}}, lp_{new}^r(m_k)+t_k\}.\\
\end{equation}

Consequently, the evaluation of $T$ in Formula (\ref{eq:STime}) for all the possible $O(n^3)$ \textit{insertion points} can be completed in $O(n^3)$ time.

\subsubsection{Computation of Communication Costs $CC$}

For an edge $(m_i,m_j)\in E_{TG}$, the communication cost between $m_i$ and $m_j$, $CC(m_i,m_j)$,  can be evaluated as follows:
\begin{equation}
\begin{split}
&CC_{m_i,m_j} =\\& w_{{i}, {j}}\cdot \{\alpha_d\cdot(|x_{{i}}-x_{{j}}|+|y_{{i}}-y_{{j}}|)+ \beta_t \cdot(bt_j - et_i)\}
\end{split}
\end{equation}
where $w_{{i},{j}}$ is the communication requirement between $m_i$ and $m_j$, and ($x_{{i}},y_{{i}}$) and ($x_{{j}},y_{{j}}$) are respectively the coordinates of $m_i$ and $m_j$ on the FPGA chip.
$et_i$ ($= bt_i + t_i$) is the ending execution time of $m_i$.

If the two task modules span multiple time layers, we project the two task modules onto one time layer and calculate the Manhattan distance. In the experiments, the parameters $\alpha_d$ and $\beta_t$ are set based on the temporal and spatial dimensions:
1) If $m_i$ and $m_j$ are partitioned into the same time layer, we set $\alpha_d$ and $\beta_t$ to 1 and 0, respectively;
2) If $m_i$ and $m_j$ are partitioned into different time layers in a DRR, then $\alpha_d$ and $\beta_t$ will be set to 1 and 1.5, respectively;
3) If $m_i$ and $m_j$ are partitioned into different DRRs, $\alpha_d$ and $\beta_t$ will be set to 3 and 1.5, respectively.

The communication cost $CC_{m_k}$ can be calculated as follows.
\begin{small}
\begin{equation}\label{equ:comm-cost}
CC_{m_k}= \sum_{(m_i,m_k)\in E_{TG} } CC(m_i,m_k) + \sum_{(m_k,m_i)\in E_{TG}} CC(m_k,m_i).
\end{equation}
\end{small}

Thus, the total communication costs $CC$ can be evaluated as follows:
$CC= \sum_{m_k\in V_{TG} } CC_{m_k}$.

\subsubsection{Computation of Area Costs}
\label{sec:compute_ac}
\
Let $Row_0$ and $Col_0$, respectively, be the {number of rows and the number of columns} of the CLB array available in the FPGA chip. The area cost $AC$ is calculated using a method similar to that in \cite{chen2008fixed}:
\begin{equation}
\label{equ:fixedfp}
\begin{split}
AC = &E_{Row}+E_{Col}\cdot \lambda+C_1\cdot max(E_{Row},E_{Col}\cdot \lambda),
\end{split}
\end{equation}
where $E_{Col}=max(Col-Col_0,0)$ and $E_{Row}=max(Row-Row_0,0)$ are respectively the
excessive columns and rows required by the current solution, $\lambda=Row_0/Col_0$, and $C_1$ is a
user-defined constant.

The $Col$ and $Row$ are evaluated in constant time by a method similar to that in \cite{chen2008fixed} and \cite{chen2010multi}. Considering a feasible insertion point $(p, q, r, tl^i_j)$, we can first calculate the row and column within the DRR using a method similar to \cite{chen2010multi} in amortized constant time because the time layers in the DRR are represented by a multi-layer sequence pair. After the dimensions of the DRR are obtained, we can use a similar method to that in \cite{chen2008fixed} to calculate the total rows and total columns required by inserting the removed task module into the insertion point $(p, q, r, tl^i_j)$ in amortized constant time. 

\vspace{-1em}
\subsection{Complexity Analysis}\label{sec:complex}

In this subsection, we analyze the complexity of an IAR perturbation on a $P$-$ST$.
For an insertion point $(p,q,r, tl^i_j)$, if conditions for both \textit{type-1} and \textit{type-3} in Section \ref{subsec:fip} are satisfied, there will be at most five possible time layer candidates: one type-1 insertion point, where $tl^i_j$ is a new created time layer in a newly-created DRR; two type-2 insertion points, where $tl^i_j$ is a newly-created time layer in an existing DRR; two type-3 insertion points,  where $tl^i_j$ is an existing time layer. Consequently, there are $O(n^3)$ possible \emph{insertion points} for a removed task module.

The feasibility of an insertion point can be judged in amortized constant time, as both the minimum lifetime in Corollary \ref{cor:lifetime-min} and the lifetime constraint in Corollary \ref{cor:lifetime-mk} can be computed previously in $O(n^2)$ time.

For each \emph{insertion point}, the cost function shown in Equation \ref{equ:evl} can be calculated in $O(n)$, where the area cost ($AC$) and the \textit{schedule length} $T$ can be evaluated in amortized constant time, and the complexity of computing communication costs is $O(n)$.
Consequently, the evaluation of $O(n^3)$ insertion points can be performed in $O(n^4)$, where $n$ is the number of task modules.

\section{Experiments and Results}
\label{sec:experiments}
\subsection{Experimental Setup}
\label{sec:exp-setup}
The proposed method has been implemented in C-language on a Linux 64-bit workstation (Intel 2.0 GHz, 62 GB RAM).
The input consists of a set of tasks with dependencies given in a task graph ($TG$), the resource requirements, configuration time $ct_i$, and execution time $et_i$ of each task module $m_i$.
The benchmarks are constructed by combining the task graphs generated by Task Graphs For Free (TGFF) \cite{dick1998tgff} and the standard floorplanning benchmark, GSRC suites \cite{GSRC}.
The dimensions (width and height) of task modules are from GSRC benchmarks and the width and the height of a task module respectively define the number of CLB columns and the number of CLB rows on an FPGA chip.
The task dependencies are generated by TGFF. The execution times of task modules and the communication requirements between task modules are randomly generated.
Note that we are considering only the allocation of CLB resources.
\revise{We also generate two benchmarks from a popular convolutional neural network model, AlexNet \cite{krizhevsky2012imagenet}.
One  is \textit{AN\_Part1}, which includes two convolutional layers and one pooling layer of the model.
The other is  \textit{AN\_Part2}, which includes three convolutional layers and two pooling layers of the model.
The task module in convolutional layers performs a convolution operation over a feature map with a specified convolutional kernel and the task module in pooling layers performs a pooling operation over all the output feature maps of a convolutional layer.
The task modules are stipulated to consume the least hardware resources, and the execution times are estimated based on a frequency of 200 MHz.}

Table~\ref{tab:bmark} lists the benchmark parameters used in our experiments.
Each \revise{randomly generated} benchmark has three different implementations, which have the same number of task modules but different task dependencies.
\#V and \#E are the number of vertexes and edges in task graph $TG$, respectively.
The columns $VWR$ and $EWR$ are the range of random values for execution time of tasks and communications between tasks, respectively.
The column $CPT$ shows the longest paths of the task graphs and the vertexes are weighted by the execution times.

\vspace{0.5em}
\begin{table}[htbp]
	\centering
	\caption{Benchmark Information}
	\addtolength{\tabcolsep}{-1.5pt}
	\begin{tabular}{c|c|c|c|c|c|c}
		\hline
		\hline
		\multirow{2}{*}{Bench.} & \multirow{2}{*}{imp.} & \multicolumn{5}{|c}{TG}\\
		\cline{3-7}
		& & \#V & \#E& VWR $(ms)$& EWR & CPT $(ms)$\\
		\hline
	    \multirow{3}{*}{t10}	
		& {1}	&  \multirow{3}{*}{{10}}  & {10}    & {(40,55)}    & {(20,30)}   & {259.2}   \\
		& {2}	&						              & {12}    & {(40,55)}   & {(20,30)}  & {359.1}   \\
		& {3}	&	  					              & {8}    & {(40,55)}    & {(20,30)}   & {151.1}   \\
		\hline
	    \multirow{3}{*}{{t30}}	
		& {1}	&  \multirow{3}{*}{{30}}  & {51}    & {(30,350)}    & {(60,610)}   & {1450.8}   \\
		& {2}	&						& {72}    & {(40,60)}   & {(20,30)}  & {786.3}   \\
		& {3}	&	  					& {71}    & {(40,60)}    & {(20,30)}   & {727.6}   \\
		\hline
		\multirow{3}{*}{t50}	
		& 1	&  \multirow{3}{*}{50}  & 33    & (40,60)    & (20,30)   & 244.0   \\
		& 2	&						& 51    & (20,180)   & (50,350)  & 464.9   \\
		& 3	&	  					& 78    & (40,60)    & (20,30)   & 595.9   \\
		\hline
		\multirow{3}{*}{t100}
		& 1	& \multirow{3}{*}{100}  & 110   & (20,180)   & (50,350)  & 703.1    \\
		& 2	&     					& 134	& (20,180) 	 & (50,350)  & 1265.6   \\
		& 3	&	  					& 147	& (20,180)	 & (50,350)  & 580.5    \\
		\hline
		\multirow{3}{*}{t200}	
		& 1	& \multirow{3}{*}{200}  & 312   & (10,390)   & (30,770)  & 3556.2   \\
		& 2	&	  					& 327	& (40,60)	 & (20,30)   & 436.4    \\
		& 3	&	 					& 403	& (10,390)	 & (30,770)  & 1482    \\
		\hline
		\multirow{3}{*}{t300}	
		& 1	& \multirow{3}{*}{300}  & 416   & (20,180)   & (50,350)  & 943.4   \\
		& 2	&	  					& 443	& (20,180)	 & (50,350)  & 1051.5   \\
		& 3	&	  					& 735	& (20,180)	 & (50,350)  & 2814.7  \\
		\hline

		\multicolumn{2}{c|}{{AN\_Part1}}	& {353}  & {352}  & {(5.6, 51)}   & {(5.4, 11.8)}  & {89.6}   \\
        \hline
		\multicolumn{2}{c|}{{AN\_Part2}}	& {738}  & {992}  & {(3.5, 51)}	& {(2.7, 11.8)}  & {104.4}   \\
		\hline
	\end{tabular}
	\label{tab:bmark}
\end{table}

We take one of the widely used Xilinx Virtex 7 series FPGA chips, XC7VX485T, as the target chip. There are about 37,950 CLBs and the ratio of rows to columns is 3:1. Therefore, the CLB array in XC7VX485T has approximately {350} rows and {117} columns.
Configuring all resources of XC7VX485T requires 50.7 $ms$ through the interface ICAP with the maximum bandwidth of 3.2 Gb/s \cite{vivado-pr-ug2017}.
Thus, we consider the time overhead of reconfiguring one CLB to be 0.0013 $ms$, and the configuration time span of a task module is proportional to the module area because the configuration time is proportional to the synthesized bitstream of a design.

In the experiments, the sum of the three coefficients in Formula (\ref{equ:evl}) is set to one and the cost values are normalized. To avoid violating the resource constraints in the final solutions, the coefficient of area cost, $\alpha$, is the dominant factor and is set to around $0.8$ to ensure almost 100\% success rate. The coefficients of \textit{schedule length} and communication costs, $\beta$ and $\gamma$, are respectively set to $0.15$ and $0.05$.  We can make a trade-off between the schedule length and the communication cost by changing $\beta$ and $\gamma$ because the area cost computed by Formula (\ref{equ:fixedfp}) will be zero if the resource constraint is satisfied. The initial solution is generated randomly and each task module is partitioned into an individual time layer in a common DRR and the configuration order is a topological ordering of the task graph. The starting temperature, the ending temperature, and the cooling ratio of the simulated annealing are respectively set to 2000, 0.01 and 0.98. The iteration number in each temperature is set to 50 for the benchmarks with less than 50 tasks. For other benchmarks, the iteration number is increased slightly along with the increasing number of task modules.

\subsection{Results and Analysis}
\label{sec:exp-int-psf}
\begin{table*}[htbp]
	\centering
\addtolength{\tabcolsep}{-2.3pt}
	\caption{Results of the proposed algorithm}
	\begin{tabular}{c|c|c|c||c|c|c|c|c||c|c|c|c|c}
		\hline
		\multirow{2}{*}{Benchmark} & \multirow{2}{*}{imp}& \multicolumn{2}{|c}{ $ILP$} & \multicolumn{5}{|c}{$TP\_PSF$} &\multicolumn{5}{|c}{$Int\_PSF$}\\	
		\cline{3-14}
		& &$T$(ms) & RunT$(s)$ & \#succ&$T$ (ms) &  $N$ & $CC$ & RunT$(s)$ &\#succ&$T$(ms) &  $N$ & $CC$ & RunT$(s)$\\
		
		\hline
		\multirow{3}{*}{t50} 	
		& 1& 299.45* & 3600  & {40\%} & {462.7} & {3.8}  &{325436}  &{21.1}  & {100\%}   & {456.9}  & {3.3}	& {216047}  & {27.5}  \\
		& 2& 468.82 & 12.2  & {20\%} & {688.2} & {4.1}  &{3982657} &{23.5}  & {100\%}   & {705.7}  & {3.9}	& {3644317} & {29.1}  \\
		& 3& 604.90 & 2.1   & {80\%} & {616.5} & {4.4}  &{610452}  &{25.1}  & {100\%}   & {618.9}  & {3.8}	& {589759}  & {29.8}  \\
		\hline
		\multirow{3}{*}{t100}	
		& 1& 706.22  & 23.5  & {60\%} & {712.3}  & {9.7}  &{6428845}  &{108.6} & {100\%}   & {715.22}  & {9.3}	& {6443798} & {102.2}  \\
		& 2& 1266.29 & 27.6  & {100\%}& {1266.3} & {9.6}  &{7665841}  &{135.3} & {100\%}   & {1266.29} & {9.4}	& {7629823} & {113.4} \\
		& 3& 581.33  & 64.9 & {80\%} & {611.1}  & {9.6}  &{13724146}  &{112.4} & {100\%}   & {603.85}  & {9.3}	& {13664719}& {106.5} \\
		\hline
		\multirow{3}{*}{t200}	
		& 1& 3557.37 & 16.8 & {100\%}& {3557.4} &{15.8}  &{53624898} &{528.5} & {100\%}  & {3557.37} & {16.9}	& {54086203}& {615.9} \\
		& 2&  436.92 & 386.4 & {100\%}& {436.92}  &{20.5}  &{3043017}  &{510.2} & {100\%}  & {436.92}  & {21.6}	& {3091106} & {532.5} \\
		& 3& 1483.24 & 772.42 & {70\%}& {1501.1}  &{21.9}  &{87560078} &{532.1} & {100\%}  & {1489.68} & {21.2}	& {87362210}& {568.4} \\
		\hline
		\multirow{3}{*}{t300}	
		& 1& 1120.05* & 3600  & {40\%}& {1110.3}  &{17.4}  & {37462179} &{1450.8} & {100\%}  & {1108.97} & {17.3}	& {36686354}& {1633.9}    \\
		& 2& 1379.49*& 3600  & {30\%}& {1135.6}  &{16.9}  & {41404278} &{1346.9} & {100\%}  & {1126.10} & {16.9}	& {41972811}& {1590.3}   \\
		& 3& 2815.2 & 2610.7 & {100\%}& {2815.2} &{19.0}  & {80345041} &{1390.7} & {100\%}  & {2815.21} & {17.8}	& {79287974}& {1692.8}    \\
		\hline
\multicolumn{2}{c|}{{AN\_Part1}}& {455.87*} & {3600}  & {100\%}& {542.1}  &{15.9}  & {1230864} &{1426.9} & {100\%}  & {539.3} & {21.5}	& {1199978}& {1113.1}    \\
        \hline
		\multicolumn{2}{c|}{{AN\_Part2}} & {NF} & {3600}  & {100\%}& {1214.25}  & {24.0}  & {2980086} & {6693.2} & {100\%}  & {1023.3} & {26.5}& {2686430}& {6642.1}   \\
		\hline		\hline
		\multicolumn{2}{c|}{Cmp.}& -&  -& {72.8\%} & 1 & 1 & 1 & - &{100\%} & {-1.2\%}& {1.03\%} & {-0.53\%}& {- }\\
		\hline
	\end{tabular}
	\label{tab:results}
\end{table*}

Table~\ref{tab:results} shows the experimental results. The proposed integrated optimization framework is called \textit{Int\_PSF}.
We execute the proposed method 10 times independently for each benchmark, and list the average results.
The columns in Table~\ref{tab:results} are organized as follows:
$T$ is the \textit{schedule length} of each design, which corresponds to the longest paths in the RCGs.
$RunT$ is the run-time of the optimization framework.
$\#succ$ is the success rate of floorplanning.
$N$ is the number of DRRs.
$CC$ indicates the communication costs calculated based on the temporal and spatial dimensions.

As a baseline situation, we solve the simplified scheduling problem, where the hardware resources are considered unlimited and every task module occupies an individual DRR, using an integer linear programming (ILP) formulation similar to that in \cite{deiana2015multiobjective}. The obtained $T$ indicates the \textit{schedule length} in the case when the configuration times are maximally hidden. In the experiment, $Gurobi$ \cite{gurobi} is used as the ILP solver.
The column $ILP$ shows the results, where the $'*'$ means the solutions are incumbent within one hour.
The results demonstrate that $Int\_PSF$ can effectively hide the reconfiguration time overhead in the dependency dominated task graphs (the designs having long CPT) under the resource constraints.

To explore the effectiveness of the proposed integrated optimization framework, we perform a two-phases approach ($TP\_PSF$) for partitioning, scheduling, and floorplanning of task modules.
In the first phase, we evaluate the hardware resources for a time layer using the sum of the task module area instead of calculating a floorplan of the task modules. Consequently, the order of task modules within a time layer (in PS and QS) makes no sense, but the partitioning of task modules and scheduling of the time layers are solved in an integrated optimization framework. In the second phase, a simulated annealing-based search is used for placing the task modules and the DRRs. The initial floorplan of DRRs and the initial floorplan of time layers are generated randomly. In the simulated annealing, the IAR perturbation is adopted for a DRR (as a whole) or a task module, and the task modules are removed and inserted only within a time layer to keep the partitioning unchanged.
As shown in Table~\ref{tab:results}, $Int\_PSF$ achieves a success rate of 100\% whereas the two-phase method  $TP\_PSF$ achieves only a success rate of {72.8\%} in the case when the \textit{schedule length} and $CC$ are almost the same.

In benchmark t50-2, the $T$ obtained by the proposed method is obviously higher than the baseline situation because t50-2 has high parallelism, whereas FPGA hardware resources constrain the parallel executions of the tasks.
Table \ref{tab:results-cmpresource} shows the detailed experimental results on the relationships between FPGA resources and performances for applications with different degrees of parallelism.

According to our experimental results, the communication costs $CC$ can be reduced by {33.18}\% on average by considering the communication costs in the optimization framework. The detailed data are listed in the supplementary material (Table I).

To evaluate the impacts of FPGA resources on the \textit{schedule length} of designs, we perform the experiments for all test benchmarks under different FPGA resource constraints, which are set as 3/4x, 1.0x, 3/2x, and 2.0x of the targeted FPGA architecture (37,950 CLBs).
We execute $Int\_PSF$ 10 times independently for each benchmark, and show the average results in Table \ref{tab:results-cmpresource}.
The column $Resource$ represents the amount of resources for the target FPGA architecture.
As shown in Table \ref{tab:results-cmpresource}, for all the benchmark circuits, with increasing FPGA resources, the trends of
\textit{schedule length} $T$ and communication costs $CC$ decrease to be gentle.
On the one hand, DRRs can be executed in parallel and configured independently, thus the configuration latency can be effectively hidden in the executions of tasks. With increased FPGA resources, the number of DRRs $N$ is increasing overall, which will maximize the parallel execution of tasks and increase the possibility of hiding the configuration of tasks. On the other hand, the \textit{schedule length} should be greater than that in the baseline situation shown in Table \ref{tab:bmark}. For the benchmark t200-1, which involves a long CPT and fewer data dependencies, the \textit{schedule length} $T$ remains the same with increasing FPGA resources and is close to the length of the corresponding CPT because the configuration latency is effectively hidden.
Furthermore, $Int\_PSF$ achieves 100\% success rate for the different FPGA resource constraints, which demonstrates the effectiveness of the method.

\begin{table*}[htbp]
{
	\centering
    \addtolength{\tabcolsep}{3pt}
	\caption{Impacts of FPGA resources on $Int\_PSF$}
	\begin{tabular}{c|c|c||c|c|c||c|c|c||c|c|c}
		\hline
      & & & \multicolumn{3}{c|}{t100} &\multicolumn{3}{|c|}{t200}&\multicolumn{3}{|c}{t300}\\
		\hline
		 imp. & Resource& \#succ & $N$ &$T$(ms) & $CC$ & $N$ &$T$(ms) & $CC$ & $N$ &$T$(ms) & $CC$ \\

		\hline
			                                             	 	                                              	
		 \multirow{4}{*}{1}& 3/4 & 100\%  & 5.6 & 887.45 & 7708821       	 & 12.5 & 3557.37 & 54554324           & 11.6 & 1344.19 & 41151018   \\
                           & 1.0 & 100\%  & 9.3 & 715.22 & 6443798          & 16.9 & 3557.37 & 54086203          & 17.3 & 1108.97 & 36686354    \\
                           & 3/2 & 100\%  & 14.1 & 706.22 & 5277481         & 18.9 & 3557.37 & 53948392         & 31.9 & 944.88 & 32868833    \\
                           & 2.0 & 100\%  & 14.7 & 706.22 & 5063156         & 17.4 & 3557.37 & 54165018         & 46.8 & 944.41 & 33023986    \\
        \cline{1-12}
		 \multirow{4}{*}{2}& 3/4 & 100\%  & 5.4 & 1270.60 & 7705509       	 & 13.2 & 478.45 & 3312604             & 11.6 & 1294.18 & 46898340    \\
                           & 1.0 & 100\%  & 9.4 & 1266.29 & 6443798         & 21.6 & 436.92 & 3091106           & 16.9 & 1126.10 & 41972811    \\
                           & 3/2 & 100\%  & 9.5 & 1266.29 & 5277481         & 27.2 & 436.92 & 2749060           & 31.6 & 1052.78 & 38627940    \\
                           & 2.0 & 100\%  & 9.9 & 1266.29 & 5063156         & 27.9 & 436.92 & 2747273           & 37   & 1052.78 & 38020370    \\
        \cline{1-12}
		 \multirow{4}{*}{3}& 3/4 & 100\%  & 5.3 & 804.19 & 15790269       	 &  13.1 & 1725.29 & 97307300           & 12.3 & 2841.89 & 81422950    \\
                           & 1.0 & 100\%  & 9.3 & 603.85 & 13664719         &  21.2 & 1489.68 & 87362210         & 17.8 & 2815.21 & 79287974    \\
                           & 3/2 & 100\%  & 15.3 & 581.33 & 11737755        &  30.4 & 1483.24 & 83528144         & 21.8 & 2815.21 & 78553805    \\
                           & 2.0 & 100\%  & 17.9 & 581.33 & 10975717        &  31.8 & 1483.24 & 83050648         & 22.5 & 2815.21 & 78461001    \\

		\hline
	\end{tabular}
	\label{tab:results-cmpresource}
}
\end{table*}


As discussed in Section \ref{sec:evaluation}, to reduce the complexity, we use the area sum of the task modules to underestimate the configuration span of the time layers instead of accurately computing the configuration span of the DRRs, which should be computed using the DRR area.
According to our experimental results, for the same solution, if the DRR area is used instead of summing the module area, the \textit{schedule length} is increased by 5\%.
In the optimization framework, if we use the DRR area to accurately estimate the configuration span of time layers, the obtained \textit{schedule length} is increased only by negligible 1\%. 
The possible reason is that the configurations of the time layers are well hidden in the execution of task modules. 
A detailed analysis is included in the supplementary material (Figure 1).


\subsection{Vertically Aligning DRRs to Reconfigurable Frames}
\label{sec:align-exp}
As was demonstrated in \cite{vivado-pr-ug2017}, when applying the \textit{Reset After Reconfiguration} methodology, a DRR must vertically align to reconfigurable frames (aligning vertically to clock regions) for 7-series and Zynq-7000 AP SoC devices.  When the height of the DRRs is vertically aligned to the reconfigurable frames, the height of the reconfiguration frame (50 rows of CLBs) is adopted as the measurement unit of DRR height. Table \ref{tab:results-align} shows the experimental results of with/without consideration of the aligning constraint. The results of the both cases are comparable. An example result of t50-1 is shown in the supplementary material (Figure 2).


\begin{table}[htbp]
{
	\centering
\addtolength{\tabcolsep}{-3.3pt}
	\caption{Aligning DRRs to Reconfigurable Frames}
	\begin{tabular}{c|c|c|c|c|c||c|c|c|c}
		\hline
		\multirow{2}{*}{Bench.} & \multirow{2}{*}{imp}& \multicolumn{4}{|c}{$Int\_PSF$ with aligning} &\multicolumn{4}{|c}{$Int\_PSF$}\\	
		\cline{3-10}
		& & \#succ&$T$ (ms) &  $N$ & $CC$ &\#succ&$T$(ms) &  $N$  & $CC$\\
			\hline
		\multirow{3}{*}{t10} 	
		& 1&  100\% & 273.6 & 3.0   &65125  & 100\%   & 269.8   & 3.9	& 54078  \\
		& 2&  100\%  & 370.8 & 2.5   &82022  & 100\%   & 369.6   & 2.5	& 75027  \\
		& 3&  100\% & 232.5 & 3.0   &38566  & 100\%   & 201.6   & 3.7	& 57750  \\
		\hline
		\multirow{3}{*}{t30}	
		& 1&  100\% & 1470.9  & 2.7  &7516360   & 100\%   & 1459.8  & 3.0	& 6806495  \\
		& 2&  100\% & 799.8   & 2.8  &582164    & 100\%   & 793.1   & 3.0	& 599203 \\
		& 3&  100\% & 738.7   & 3.0  &602132    & 100\%   & 734.2   & 3.0	& 615542 \\
		\hline
		\multirow{3}{*}{t50}	
		& 1&  100\%& 473.6    &4.0   &257776    & 100\%  & 456.9    & 3.3	& 216047 \\
		& 2&  100\%& 691.2    &4.6   &3826090   & 100\%  & 705.7    & 3.9	& 3644317 \\
		& 3&  90\% & 611.9    &4.7   &625038    & 100\%  & 618.9    & 3.8	& 589759 \\
		\hline
        \hline
		&-& 98.9\% & 629.2 &3.36 &1510585 & 100\%& 623.3 & 3.34& 1406468 \\
		\hline
	
		\hline
	\end{tabular}
	\label{tab:results-align}
}
\end{table}

\vspace{-1em}
\subsection{Comparison with Previous Work}
\label{sec:comp}
R. Cordone et al. \cite{cordone2009partitioning} proposed a partitioning method to extract cores (isomorphic and non-overlapping subgraphs) from the task graphs for module reuse and an integer linear programming (ILP) based method and a heuristic method for scheduling task graphs on partially dynamically reconfigurable FPGAs.
The core extraction method provides preprocessing of the task graphs and can be combined with other scheduling methods to consider module reuse.
However, it is difficult to extend the scheduling method for processing DRR partitions, while the task partitioning algorithm in Y. Jiang et al. \cite{jiang2007temporal} can be used only within a DRR.
E. A. Deiana et al. \cite{deiana2015multiobjective} proposed a mixed-integer linear programming (MILP) based scheduler for mapping and scheduling applications on partially reconfigurable FPGAs with consideration of DRRs, where only one task module is involved in each time layer, followed by floorplanning the DRRs.
Consequently,
in this study, we make a comparison with the methodology in \cite{deiana2015multiobjective}.
We adapt the ILP method in \cite{deiana2015multiobjective} to the problem in this work by skipping the module reuse, and in the proposed optimization framework, we add a constraint (\textit{one task constraint}) so that each time layer includes only one task module. As for the floorplanning of DRRs, there are no detailed descriptions in \cite{deiana2015multiobjective}, so we use the method in \cite{chen2008fixed}, which performs very well in the fixed-outline-constrained floorplanning for FPGAs and spends only several seconds for the floorplanning of 100 task modules.

Table \ref{tab:results-compareILP} shows the experimental results. \revise{The results are the average of 10 independent runs. In the \textit{ILP+Flooprlan} method, we solve the ILP model once and run the floorplanning algorithm 10 times.}
Because the ILP based method is {time-consuming}, we use the small test cases, t10, t30, and t50 (the largest test cases in \cite{deiana2015multiobjective} includes $50$ tasks), for the comparison. \revise{\textit{$'NF'$} represents that the ILP solver fails to find any feasible solution in a reasonable time.}
The results show that the proposed optimization framework achieves much higher success rates with comparable \textit{schedule lengths}.

\begin{table}[htbp]
{
	\centering
\addtolength{\tabcolsep}{-3.7pt}
	\caption{Comparison between the $ILP$+$Floorplanning$\cite{deiana2015multiobjective} and the proposed $Int\_PSF$}
	\begin{tabular}{c|c|c|c|c|c||c|c|c|c}
		\hline
		\multirow{2}{*}{Bench.} & \multirow{2}{*}{imp}& \multicolumn{4}{|c}{$ILP$+$Floorplan$\cite{deiana2015multiobjective}} &\multicolumn{4}{|c}{$Int\_PSF$(one task constraint)}\\	
		\cline{3-10}
		& & \#succ&$T$ (ms) &  $N$ & RunT$(s)$ &\#succ&$T$(ms) &  $N$  & RunT$(s)$\\
			\hline
		\multirow{3}{*}{t10} 	
		& 1&  100\% & 269.8 & 4.0   &0.64  & 100\%   & 269.8   & 4.0	& 1.56  \\
		& 2&  90\%  & 369.7 & 4.0   &0.54  & 100\%   & 369.7   & 3.0	& 1.64  \\
		& 3&  100\% & 176.2 & 4.0   &2.24  & 100\%   & 202.6   & 4.0	& 1.47  \\
		\hline
		\multirow{3}{*}{t30}	
		& 1&  50\% & 1459.8  & 6.0  &7.1   & 100\%   & 1460.1  & 4.6	& 6.82  \\
		& 2&  80\% & 792.8   & 6.0  &7.9   & 100\%   & 794.4   & 5.1	& 6.55 \\
		& 3&  30\% & 734.2   & 5.0  &17.8  & 100\%   & 739.1   & 4.7	& 6.47 \\
		\hline
		\multirow{3}{*}{t50}	
		& 1&  NF& NF &NF  &$>$10000          & 100\%  & 487.6    & 6.7	& 16.08 \\
		& 2&  NF& NF &NF  &$>$10000          & 100\%  & 733.3    & 6.6	& 17.17 \\
		& 3&  30\%& 605  &8.0  &1085       & 100\%  & 623.4    & 6.8	& 17.61 \\
		\hline
        \hline
		&-& 68.6\% & 630 & - &- & 100\%& 637(+1.1\%) & -& - \\
		\hline
	
		\hline
	\end{tabular}
	\label{tab:results-compareILP}
}
\end{table}

\section{Conclusions}\label{sec:conclusion}
In this paper, we proposed an integrated optimization framework for partitioning, scheduling, and floorplanning partially dynamically reconfigurable FPGAs, where the partitioned sequence triple $P$-$ST(PS, QS, RS)$ was proposed to represent the partitions, schedule, and floorplan of the task modules, and a sufficient and necessary condition is given for the feasibility of $P$-$ST$ considering the scheduling problem. An elaborated method was proposed to generate new solutions by simultaneously perturbing the partition, schedule, and floorplan.
Based on the proposed optimization framework, we integrated the exploration of spatial and temporal design space to search the optimal solutions of partitioning, scheduling, and floorplanning. Experimental results demonstrated the effectiveness of the proposed framework.
In future work, we will further consider the reuse of task modules, variable dimensions for task modules, and integration of the allocation of RAM and DSP resources.





\begin{IEEEbiography}[{\includegraphics[height=1.24in,clip,keepaspectratio]{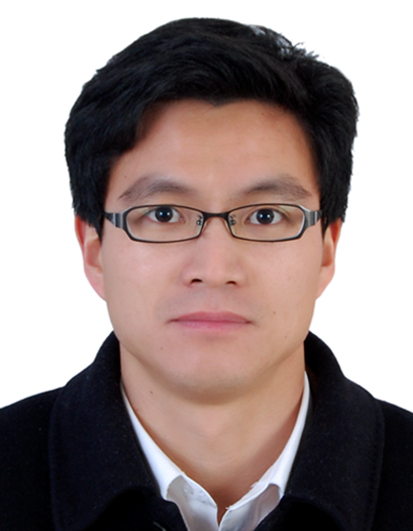}}]{Song Chen}
received his B.S. degree in computer science from Xi'an Jiaotong University, China in 2000. Subsequently, he obtained a Ph.D. degree in computer science from Tsinghua University, China in 2005.
He served at the Graduate School of Information, Production and Systems, Waseda University, Japan, as a Research Associate from August 2005 to March 2009 and as an Assistant Professor from April 2009 to August 2012.
He is currently an Associate Professor in the Department of Electronic Science and Technology, University of Science and Technology of China (USTC).
His research interests include several aspects of VLSI design automation, on-chip communication system, and computer-aided design for emerging technologies.
Dr.~Chen is a member of IEEE and IEICE.
\end{IEEEbiography}
\begin{IEEEbiography}[{\includegraphics[height=1.24in,clip,keepaspectratio]{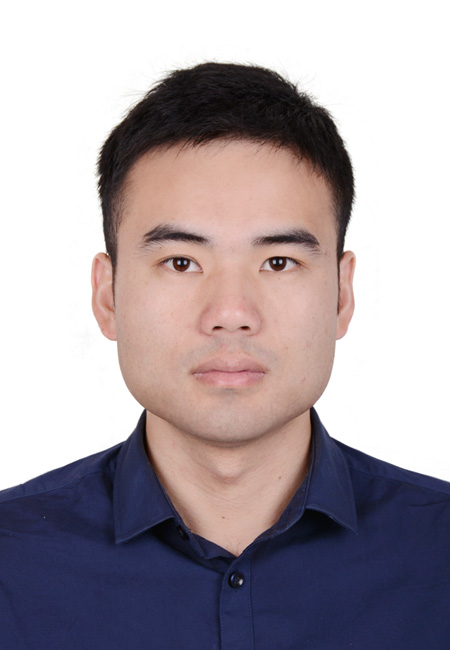}}]{Jinglei Huang}
received the B.E. degree in electronic science and technology from AnHui University, Hefei, China in 2013, and obtained a Ph.D. degree in electronic science and technology from  University of Science and Technology of China (USTC) in 2018.
He is currently an Engineer in the State Key Laboratory of Air Traffic Management System and Technology, the 28th Research Institute of China Electronics Technology Group Corporation. 
His research interests include network-on-chip synthesis and air traffic management.
\end{IEEEbiography}

\vspace{0.5em}
\textbf{Xiaodong Xu}, Photograph and biography not available at the time of publication.

\vspace{0.5em}
\textbf{Bo Ding}, Photograph and biography not available at the time of publication.
\begin{IEEEbiography}[{\includegraphics[height=1.24in,clip,keepaspectratio]{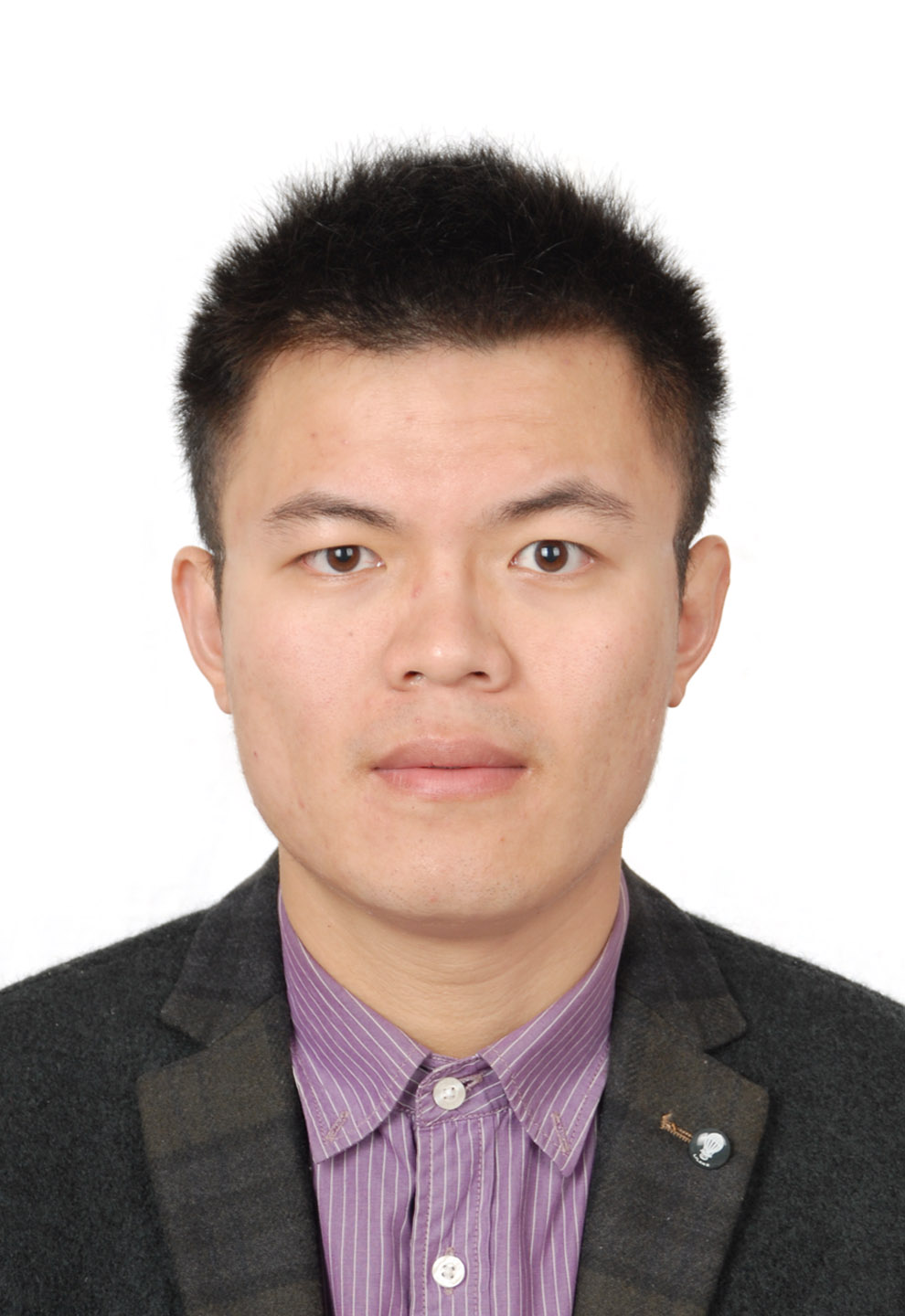}}]{Qi Xu}
received the B.E. degree in microelectronics from AnHui University in 2012 and the Ph.D degree in electronic science and technology from University of Science and Technology of China in 2018. He is currently a lecturer in the School of Electronic Science and Applied Physics, Hefei University of Technology.

His research interests include physical design automation and design for reliability for 3-D integrated circuits.
\end{IEEEbiography}

\end{document}

%% file: ieee-compact.tex
\usepackage{algpseudocode}                                 
\usepackage{algorithm}
\usepackage{graphicx}                                      
\usepackage{amsmath}
\usepackage{amssymb}
\usepackage{amsfonts}
\usepackage{booktabs}
\usepackage{multirow}
\usepackage[subrefformat=parens,farskip=0pt,justification=centering]{subfig}
\usepackage{color}
\usepackage{cite}                                          
\usepackage{comment}                                       
\usepackage{soul}                                          
\soulregister\cite7
\soulregister\ref7
\soulregister\pageref7
\usepackage{amsthm}
\usepackage{etoolbox}                                      
\usepackage{url}
\usepackage{nth}                                           
\usepackage{bm}                                            
\usepackage{balance}
\usepackage{threeparttable}
\usepackage[bookmarks=false]{hyperref}                     %
\hypersetup{
    colorlinks = true,
    citecolor  = blue,
    linkcolor  = blue,
    urlcolor   = blue,
}
\usepackage{filecontents}                                  
\usepackage{pgfplots}
\usepackage{pgfplotstable}
\pgfplotsset{compat=newest}

\usepackage[top=0.8in,bottom=0.8in,left=0.6in,right=0.6in]{geometry}
\algrenewcommand\textproc{\texttt}

\makeatletter
\let\OldStatex\Statex
\renewcommand{\Statex}[1][3]{%
  \setlength\@tempdima{\algorithmicindent}%
  \OldStatex\hskip\dimexpr#1\@tempdima\relax
}
\makeatother

\RequirePackage[normalem]{ulem} 
\RequirePackage{color}\definecolor{RED}{rgb}{1,0,0}\definecolor{BLUE}{rgb}{0,0,1} 